\documentstyle[12pt,epsf]{article}
\def\del        {  \partial  }
\def\half       {  {1\over 2}  }

\def\defint#1#2 {  \int_{#1}^{#2}  }
\def\rootof#1   {  \left( #1 \right)^{1/2}  } 
\def\deldel#1   {  {\partial\over \partial #1}  }

\def\abs#1      {  \vert #1 \vert  }
\def\ie         {  {\it i.e.}      }
\def\evalat#1   {  \left\vert_{#1} \right. } 
\def\e          { {\rm e}  }

\def\lsim    {\lower .65ex \hbox{\ $\stackrel{<}{\sim}$\ } }
\def\gsim    {\lower .65ex \hbox{\ $\stackrel{>}{\sim}$\ } }
\def\calA       { {\cal A} }

\def\calM       { {\cal M} }
\def\calN      { {\cal N} }
\def\calO       { {\cal O} }

\def\vecii#1#2      {  \left(\begin{array}{c}#1\\#2\end{array}\right)  }
\def\veciii#1#2#3   {  \left(\begin{array}{c}#1\\#2\\#3\end{array}\right)  }
\def\veciv#1#2#3#4  {  \left(\begin{array}{c}#1\\#2\\#3\\#4
                                 \end{array}\right)  }
\def\vecv#1#2#3#4#5 {  \left(\begin{array}{c}#1\\#2\\#3\\#4\\#5
                                 \end{array}\right)  }

\def\matrixii#1#2#3#4            {  \left(\begin{array}{cc}#1&#2\\#3&#4
                                       \end{array}\right) }
\def\matrixiii#1#2#3#4#5#6#7#8#9 {  \left(\begin{array}{ccc}#1&#2&#3\\
                                     #4&#5&#6\\#7&#8&#9\end{array}\right)  }
\def\mativ#1#2#3#4               {  \left(\begin{array}{cccc}
                                       #1\\#2\\#3\\#4\end{array}\right) }
\def\matv#1#2#3#4#5              {  \left(\begin{array}{ccccc}
                                     #1\\#2\\#3\\#4\\#5\end{array}\right)  }
\def\eqabegin         {  \begin{eqnarray}  }
\def\eqaend           {  \end{eqnarray}  }
\def\nn               {  \nonumber  }
\def\bracetwo#1#2     {  \left\{ \begin{array}{l} #1 \\ #2 \end{array}
                         \right.  }
\def\bracetwocases#1#2#3#4  {   \left\{ \begin{array}{ll} #1 &
                                 \qquad #2 \\
                                 #3 & \qquad #4 \end{array} \right.  }
\def\bracebegin#1     {  \left\{ \begin{array}{#1}   }
\def\braceend         {  \end{array}\right.   }
\def\parn              {  \par\noindent }
\def\parbigskip        {  \par\bigskip  }
\def\parmedskip        {  \par\medskip  }
\def\parsmallskip      {  \par\smallskip  }
\def\parbigskipn        {  \par\bigskip\noindent  }

\def\parsmallskipn      {  \par\smallskip\noindent  }
\def\parag#1           {\paragraph{#1} \mbox{ }\parmedskip\noindent}
\def\boxit#1#2      {  \vbox{\hrule\hbox{ \hskip -4.1pt \vrule\kern3pt 
                       \vbox
                    {  \hsize #1 \strut\kern3pt #2 \kern3pt\strut  }
                       \kern3pt  \vrule} \hrule  } }
\def\centerbox#1#2  {  \mbox{  }\par\bigskip  \hfil \boxit{#1}{#2} \hfil
                       \par\bigskip\noindent }
\def\rightbox#1#2   {  \hfill\boxit{#1}{#2}  }
\def\leftbox#1#2    {  \boxit{#1}{#2}  }
\def\fullbox#1      {  \boxit{\textwidth}{#1}  }
\def\trianglemap#1#2#3#4#5#6  {   {\large $$ \begin{array}{rcl} #1\!\!\!
                                  &{\stackrel{{\scriptstyle #2}}{
                              \longrightarrow   }}&\!\!\!  #3 \\ 
                            { } & {\scriptstyle #4}\!\!\!\searrow \quad
                                \swarrow \!\!\!{\scriptstyle #5}& { } \\
                                  { } & #6 & { } \end{array} $$ }    }
\def\squaremap#1#2#3#4#5#6#7#8    { {\large $$ \begin{array}{ccc}#1 &
                   \stackrel{{\scriptstyle #2}}{\longrightarrow} & #3 \\
                     {\scriptstyle #4}\!\downarrow & { } & \downarrow \!
                     {\scriptstyle #5}\\ #6 &\!\!
                      \longrightarrow_{{ }_{\!\!\!\!\!\!\!\!\!\!\!
                      {\scriptstyle #7}}}    &#8 \end{array} $$ }   }
\def\righttrianglemap#1#2#3#4#5#6  {  {\large $$ \begin{array}{rcl}
                 #1\!\! & \stackrel{{\scriptstyle #2}}{\longrightarrow} 
                      & #3 \\  { }&\!\!{\scriptstyle #4}\!\!\searrow
                      & \downarrow \!\!{\scriptstyle #5}\\
                      { }&{ }& #6 \end{array} $$ }   }
\def\rightfigspacebegin  {  \par\noindent\begin{minipage}[t]{10cm}  }
\def\rightfigspaceend    {  \end{minipage}\par\noindent  }
\def\leftfigspacebegin   {  \par\noindent
                             \hspace*{10cm}\begin{minipage}[t]{6cm} }
\def\leftfigspaceend     {  \end{minipage}\par\noindent  }
\def\titleandfile#1#2   {  \begin{center}{\Large\bf #1}\end{center}
                            \par\begin{flushright} #2 \end{flushright}  }
\def\msection#1      {  \begin{center} \section{#1} \end{center}   }
\def\nsection#1      {  \let\boldface\bf \def\bf{} \section{#1}
                           \let\bf\boldface   }
\def\mnsection#1     {  \begin{center} \nsection{#1} \end{center}  }
\def\capsection#1    {  \let\boldface\bf \def\bf{\sc} \section{#1}
                           \let\bf\boldface   }
\def\mcapsection#1   {  \begin{center} \capsection{#1} \end{center} }
\def\sectionnumbering { \setcounter{equation}{0}
         \renewcommand{\theequation}{\arabic{section}.\arabic{equation}}}
\newcommand{\nullify}[1]{}
\def\sindent{\mbox{}\indent}
\def\period{\, .}
\def\comma{\, ,}
\def\overi{{1\over i}}
\def\ff{\gamma^2}  % front factor
\def\xplus{{x^+}}
\def\xminus{{x^-}}
\def\yplus{{y^+}}

\def\delplus{\del_+}
\def\delminus{\del_-}

\def\ket#1{\mid #1 >}
\def\bra#1{< #1 \mid }

\def\calA{{\cal A}}

\def\downvacket{\mid 0 >_\downarrow}
\def\pdownvac{\mid \vec{P} >_\downarrow}

\def\psmearvac{\mid \tilde{P} >} %sato
\def\psmearnorm{ < \tilde{P} \mid  \tilde{P} >} %sato
\def\Psizeroket{\mid \Psi_0 >}%sato
\def\weight{W(p^+)}%sato
        \def\Ldl{L^{dL}}

\def\Ltot{L^{tot}}

\def\Pplus{P^+} 

\def\vecpf{\vec{p}_f}

\def\vecalf{\vec{\alpha^f}}

\def\vecA{\vec{A}}
\def\dhat{\hat{d}}

\def\alplus{\alpha^+} \def\alminus{\alpha^-}

\def\pplus{p^+} \def\pminus{p^-}
\def\qplus{q^+} \def\qminus{q^-} %added 8/16

\def\Kplus{K^+}  \def\Kminus{K^-}

\def\invNhat{\hat{N}^{-1}_{dLg}}

\def\psit{\tilde{\psi}}
\def\psip{\psi^+}

\def\chip{\chi^+}
\def\chitp{\tilde{\chi}^+}

\def\sqfp{\sqrt{4\pi}}

\def\etap{\eta^+}

\def\etazp{\eta_0^+}

\def\pe{p_\eta}
\def\qe{q_\eta}
\def\pz{p_\zeta}
\def\qz{q_\zeta}
\def\At{\tilde{A}}
\def\zetap{\zeta^+}
\def\xpint{\int_0^{2\pi} {d\xplus \over 2\pi}}
\def\ypint{\int_0^{2\pi} {d\yplus \over 2\pi}}
\def\Atil{\tilde{A}}

\def\ox{{\cal O}(x)}
\def\Psizero{\Psi_0}

\def\alf{\alpha^f}

\def\Psig{\Psi_{-1}}
\def\Psibarg{\bar{\Psi}_{-1}}
\def\Ltot{L^{tot}}
\def\Lbartot{\bar{L}^{tot}}
\def\dbar{{\bar{d}}}
\def\bbar{{\bar{b}}}

\def\calLm{{\cal L}_-}
\def\calLp{{\cal L}_+}
\def\calM{{\cal M}}
\def\Dhat{\hat{D}}

\def\gt{\tilde{\gamma}}

\def\chia{\chi_a}

\def\nchiepsi{:\chi\e^{-\psi}: }
\def\At{\tilde{A}}
\def\etap{\eta^+}

\def\psitz{{\tilde{\psi}=0}}
\def\psit{\tilde{\psi}}
\def\alf{\alpha^f}
\def\Gf{{G_f}}
\def\Omegaket{\mid \Omega >}
\def\Omegabra{< \Omega \mid }
\def\omegavac{ \Omegaket} 
\def\Fii{{}_1F_1}
\def\Fiialpha#1#2{ \Fii\left( #1;#2; -{\xi^2 \over 2a(1+x^2)}\right) }
\def\Fiia#1#2{\Fii\left( #1;#2; -{\xi^2 \over 4a}\right) }

\def\xilarge{&\stackrel{\xi\rightarrow \infty}{\sim}& }
\def\cx#1#2{ (1-x^2)^{#1} (1+x^2)^{#2} }
\def\Eplus#1{ \e^{-i #1 \qplus / \pplus }}
\def\lm{\lambda_g}
\def\kf{K_g}

\def\xiplus{\xi^+}
\def\ximinus{\xi^-}

\def\appendixnumbering#1{\setcounter{equation}{0}
 \renewcommand{\theequation}{{#1}.\arabic{equation} }}
\def\np    { Nucl. Phys. }
\def\pr    { Phys. Rev. }
\def\pl    { Phys. Lett. }

\def\jmp   { J. Math. Phys. } 
\def\papertitlepage{\baselineskip 3.5ex \thispagestyle{empty}}
\def\Title#1{\vspace{1.5cm}\begin{center}
 {\Large\bf #1} \end{center} 
\vspace{1.5cm}}
\def\Authors#1{\begin{center} {\it #1} \end{center}}
\def\Abstract{\vspace{1.5cm}\begin{center} {\large\bf Abstract} 
           \end{center} \parbigskip}
\def\Komabanumber#1#2#3{\hfill \begin{minipage}{4cm} UT-Komaba #1
              \parn #2\parn #3 \end{minipage}}
\renewcommand{\thefootnote}{\fnsymbol{footnote}}
\renewenvironment{thebibliography}{\pagebreak[3]\par\vspace{0.6em}
\begin{flushleft}{\large \bf References}\end{flushleft}
\vspace{-1.0em}

\begin{enumerate}\if@twocolumn\baselineskip=0.6em\itemsep -0.2em
\else\itemsep -0.2em\fi\labelsep 0.1em}{\end{enumerate}}
\begin{document}
%%%%%%%%%%%%%%%%%%%%%%%%%%%%
\papertitlepage
\vspace*{0cm}
\Komabanumber{93-19}{hepth@xxx/9310156}{October 1993}
%%%%%%%%%%%%%%%%%%%%%%%%%%%%
\Title{Space-time Geometry \\ \vskip 1.5ex in  \\ 
\vskip 1.5ex Exactly Solvable  Quantum Dilaton Gravity} 
\Authors{{\sc Y.~Kazama
\footnote[2]{e-mail address:\quad  
kazama@tkyvax.phys.s.u-tokyo.ac.jp}
 \  and\ \ Y.~Satoh
%} \\ 
\footnote[3]{e-mail address:\quad
ysatoh@tkyvax.phys.s.u-tokyo.ac.jp} } \\
\vskip 3ex
 Institute of Physics, University of Tokyo, \\
 Komaba, Tokyo 153 Japan \\
  }
%\vspace{1.5cm}
%%%%%%%%%%%%%%%%%%%%%%%%%%%%%%%%%%%%%%%
\Abstract
%%%%%%%%%%%%%%%%%%%%%%%%%%%%%%%%%%%%%%
\baselineskip=0.7cm 
We describe in detail how one can extract space-time geometry in an 
exactly solvable model of quantum dilaton gravity of the type 
proposed by Callan, Giddings, Harvey and Strominger ( CGHS ). 
Based on our 
previous work, in which a model with 24 massless matter scalars 
was quantized rigorously in BRST operator formalism, we compute,
without approximation,  mean values of the matter stress-energy 
tensor, the inverse metric and some related quantities in a class of 
coherent physical states constructed in a specific gauge within the 
conformal gauge.
 Our states are so designed as to describe a variety of space-time 
 in which in-falling matter energy distribution produces a black hole 
 with or without naked sigularity.  In particular, we have been able 
 to produce the prototypical configuration first discovered by CGHS, 
in which a ( smeared ) matter shock wave produces a black hole without 
 naked sigularity.  
%%%%%%%%%%%%%%%%%%%%%%%%%%%%%%%%%%%%%
%\\
%\\
%PACS number(s): 04.06.+n
\newpage
\baselineskip=0.7cm
%%%%%%%%%%%%%%%%%%%%%%%%%%%%%%%%%%
\section{Introduction}
%%%%%%%%%%%%%%%%%%%%%%%%%%%%%%%%
\renewcommand{\thefootnote}{\arabic{footnote}}
\sectionnumbering
\sindent
Many of the perplexing difficulties in quantum gravity are intimately 
associated with its physical interpretation.  
This stems from the ironic circumstance that while 
 geometry is the deep key concept that captures the essence of gravity, 
our actual perception inherently hinges upon local measurements and 
hence  cannot truely be geometrical.  As a matter of fact this dilemma 
already exists in classical general relativity; no one knows how to 
describe physics in terms of a set of coordinate-independent 
numbers alone.  Of course in classical case we know a way
 to circumvent this difficulty:  We set up a 
suitable coordinate system, which we know how to interpret in 
relation to physical measurements made in our vicinity, and relying on 
this intuition we can extend our understanding to the whole of the 
space-time manifold.  In quantum theory, however, the situation is 
much more non-trivial for various reasons.  Let us list a few which will 
 be relevant.  First, the notion of a quantum state is global. 
It describes a state of the whole system at once and no \lq\lq local" 
information is stored in itself.  The second problem,  related 
to the first, is that in the most poplular formulation, where quantum 
gravity is treated as  a constrained system \cite{WDW}
, the wave functions are 
functionals of the fields and together with the lack of probability 
interpretation  no shadow of space-time physics is recognized in them. 
\parsmallskip
%%%%%%%%%%%%
As long as one stays within the approximation where one 
deals  only with  small quantum fluctuations around a prescribed 
background geometry ( possibly with some back reaction incorporated ), 
these problems essentially do not present themselves. 
However, with the recent developments of quantum gravity, especially in 
two dimensions where one now has models which are exactly solvable, 
this problem of physical interpretation has become  one of the 
central issues to be faced squarely.  The purpose of this article 
 is to discuss this problem in concrete and exact terms in a model 
 of quantum dilaton gravity of the type proposed by Callan, Giddings, 
 Harvey and Strominger \cite{CG},\cite{RS1}-\cite{RU},
 \cite{AL1}-\cite{HT}. 
 \parsmallskip
In our previous work, hereafter referred to as I \cite{HKS},
 we have rigorously
quantized  a version of such a class of models with 24 massless matter 
scalars  by developing a non-linear and non-local quantum canonical 
mapping of interacting fields into a set of free fields.  Furthermore, 
all the physical states and operators of the model have been obtained 
as BRST cohomology classes.  Technically this constitutes the exact 
solution of the model.  However, as emphasized above, solvability and 
understandability are two different concepts in quantum gravity.  
Physical states obtained in I are expressed in terms of Fourier mode 
operators of the auxilliary free fields and as they stand they do not 
yield to physical interpretations.  \parsmallskip
%%%
In order to extract the physical meaning of these abstract states, 
one must act on them by appropriate operators of physical significance 
and see the response.  Indeed this is what we must do for as simple a 
theory as that of a single quantum harmonic oscillator: A Fock state 
by itself carries no physical meaning. Only by looking at its response 
to the action of the energy operator and by computing the expectation 
values of the  coordinate and/or the momentum opeartors,  can we 
understand the  physical content of such an  abstract state.  
\parsmallskip
%%%
In gauge theories, these physics-probing operators should preferably 
be gauge invariant. The problem in the case of gravity, however, is that 
except for such an operator as the volume of the universe, 
there are few simple gauge invariant operators which we know how to 
interpret.  Conceptually, one may imagine introducing gauge invariant
 \lq\lq clocks and rulers " and try to describe the motion of particles 
 and fields in relation to these quantities.  In practice, however, 
 it is extremely difficult, if not impossible,  to construct such 
 measuring apparatus out of the fields in a given model: One is 
 free to pick certain gauge invariant quantities and declare 
them as one's reference entities, but there is no guarantee  
 that they will allow us to extract intuitively understandable 
 physics.  Although an attempt in this direction has recently been made 
 \cite{VV1,VV2}, it is not 
 clear how classical space-time picture can be reconstructed from the 
 first principle in this approach. \parsmallskip
%%%%
This brings us to the remaining alternative,{\it i.e.}, to the use of more 
 familiar operators, such as the metric, the curvature and the 
energy-momentum tensor of the matter fields, as our probe. These 
operators are obviously gauge dependent and hence in order to get 
definite responses we must fix the gauge completely.  In the BRST 
formalism  we are adopting, this corresponds to making a definite
 choice of a representative for each non-trivial cohomology class.
 What is the suitable criterion for making such a choice ? 
 It is connected to another fundamental issue, namely which matrix 
elements we should compute and how to interpret them. 
 Our point of view is the  following : Even with a lack of  
 probabilistic interpretation of the wave functions,  mean values 
 of the operators listed above in a chosen state should be related 
 to what we actually observe in a universe specified by that state. 
 In particular, if we arrange a suitable state, classical geometry 
 ( with quantum corrections )  should be recognizable in such averages. 
Among the classical solutions of the CGHS model, the most interesting 
 is the one in which a matter shock-wave produces a black hole 
 configuration. Thus we shall try to choose a cohomology class and a 
particular representative  thereof so that such a configuration is 
reproduced.  For technical reasons, we shall be able to  deal only 
 with a few of the desired operators, including the matter 
stress-energy tensor and the inverse metric $g^{\alpha\beta}$. 
Nevertheless, we shall be able to compute, without approximation, 
 the mean values of these  operators in a certain class of 
 coherent physical states and see that black holes 
 with and without naked singularities can be formed by  
 smeared shock-wave-like in-falling matter distributions.
 Although an attempt has recently been made \cite{AL2},   
 this is to our knowledge the first time that 
one can explicitly see the emergence of space-time geometry in 
 an exactly solvable model of quantum dilaton gravity containing
matter fields. \parsmallskip
%%%%%%%%
In the course of our calculation, we face the question of 
 the choice of the inner product between states, especially in 
 the space of zero-modes of the dilaton-Liouville sector
 which is generated by hermitian operators with continuous spectra. 
As was analyzed some time ago in \cite{AF1,AF2}, 
essentially two types of  inner product 
 can be consistently implemented in such a case.  One of them 
 involves indefinite metric and was later shown \cite{AF2} to 
 be relevant for the prescription of 
the \lq\lq conformal rotation" \cite{GHP} 
in the Euclidean path integral  formulation of four dimensional 
Einstein quantum gravity. 
 In the present case,  however, we find that such a choice is 
 in conflict with the requirement of reality of 
 various mean values.  Instead the
 correct choice turned out to be of the remaining type in the 
classification  of \cite{AF1}.  More details will be 
provided later. \parsmallskip
%%%%%%%%
We orgainze the rest of this article as follows: In Sect. 2, we 
 provide a brief review of the results obtained in I, in preparation 
for the subsequent sections. Expressions of the physical states 
obtained there in the BRST formalism, however, are not quite useful
 for our purposes.  Therefore in Sect. 3 we construct two different 
DDF-type representations \cite{DDF},
 which will be used in the actual calculations.
 Some techinical details concerning this construction are relegated 
 to  Appendix A.  Sect. 4 and 5 constitute the main part of our work.
In Sect.4, we first give some motivations for the class of physical 
 states we shall consider and  write down their explicit 
 forms.  Then, following a discussion of the choice of the inner 
 product, we describe the essence of the acutual computation of the 
 mean values for the operators mentioned previously.  
 Details of the exact results, which 
 are rather involved,  are listed in  Appendix B. 
 In Sect. 5, we focus on a class of particularly interesting cases 
and analyze them  in the limit where the (parameter) size of the universe 
becomes large. We shall be able to show explicitly how the presence of 
 the matter energy flux produces black hole configurations of various 
 sorts. The properties of the itegrals that appear in the analysis are 
 given in Appendix C.  
Finally, in Sect. 6, we discuss the remaining problems, including 
the difficult question of how to define and compute the S-matrix. 
The essence of  our work has been reported in \cite{KS}
 %%%%%%%%%%%%%%%%%%%%%%%%%%
%%%%%%%%%%%%%%%%%%%%%%
\section{Brief Review of the Model}
%%%%%%%%%%%%%%%%%%%%%%%%
\sectionnumbering
\sindent
We begin by giving a brief review of the model and the results 
 previously obtained in I in order to make this article reasonably 
 self-contained.  This will at the same time serve to 
define various quantities to be used in the subsequent sections.
 \parsmallskip
%%%%%%%
The classical action of our model is taken to be of CGHS form \cite{CG}, 
given by 
\eqabegin
    S &=& {1 \over \ff} \int d^2 \xi \sqrt{-g}\left\{\e^{-2{\phi}}\left[
    -4g^{\alpha\beta} \partial_\alpha\phi\partial_\beta\phi
       -R_g  +4\lambda^2  \right]\ 
      + \sum_{i=1}^N \half g^{\alpha\beta}\partial_
         \alpha f_i\partial_\beta f_i\right\} \comma
   \label{eqn:cghs}       
\eqaend
where $\phi$ is the dilaton field and  $f_i\ (i=1,\ldots, N) $ are  
N massless scalar fields representing matter degrees of freedom. 
 We shall stay throughout in Minkowski space and use the metric 
convention  such that for flat space $\eta_{\alpha\beta} = 
{\rm diag}(1,-1)$. ( Compared with the form we adopted in I, 
the signs of the terms in the bracket $\left[
\ \ \right]$ are reversed to conform to the original CGHS action. 
This leads to a minor sign change, to be indicated later, for the 
results obtained in I.)
\parsmallskip
In I,  we adopted the usual convention of setting both the speed of 
light and $\hbar$ to be unity.  In this work, in order to critically 
 examine the notion of  \lq\lq quantum corrections ", we  shall 
explicitly retain  $\hbar$ dependence after quantization. 
This in turn requires us to properly 
keep track of dimensions of various quantities. In two dimensions, 
all the fields appearing in the action are dimensionless and  
the only dimensionful quantities at the classical level are 
$\lambda$ ( the dilatonic cosmological constant ) and $1/\gamma^2$ 
factor in front. They set the fundamental length and the mass scale,
 $L_0$ and $M_0$ respectively, as
\eqabegin
 L_0 &=& {1\over \lambda}, \qquad M_0 = {\lambda \over \gamma^2} \period
\eqaend
Note that $\gamma$ has the dimension of $1/\sqrt{\hbar}$. 
\parsmallskip
In order to define all the quantities unambiguously, we take 
our universe to be spatially periodic with period $2\pi L$. It is then
convenient to introduce the dimensionless coordinates 
$x^\mu = (t, \sigma) = \xi^\mu / L $ and require that all the fields 
in the action be invariant under $\sigma \rightarrow \sigma + 2\pi $. 
When the action is rewritten in terms of $x^\mu$, it retains its form 
except with the replacement $\lambda \rightarrow \mu \equiv 
\lambda L $, where  $\mu$ is dimensionless. Later when we come to 
 the physical interpretation of the results, we will get back to
 the original variables $\xi^\mu$ and $\lambda$. \parsmallskip
Quantization of this model enforcing  conformal invariance was 
proposed by several authors \cite{AL1,BC,HA,HT}
 and we adapted the approach of Ref \cite{HT}.  In their scheme, 
 one first makes 
 a classical transformation of fields 
\eqabegin
   \Phi &\equiv& \e^{-2\phi}\comma \qquad 
    h_{\alpha\beta} \equiv \e^{2\omega}g_{\alpha\beta} \comma
\label{eqn:Phih}
\eqaend
where 
\eqabegin
\omega &=& \frac12\left(\ln\Phi - \Phi\right) \period \label{eqn:omega}
\eqaend
The action thereby takes the form proposed by Russo and Tseytlin \cite{RT}
\eqabegin
    S &=& \frac1{\ff}\int d^2x\sqrt{-h}\left[-
    h^{\alpha\beta} \del_\alpha\Phi \del_\beta\Phi
      -R_h\Phi  +4\mu^2\e^\Phi + \sum_i \frac12 h^{\alpha\beta}\partial_
         \alpha f_i\partial_\beta f_i\right] \comma
   \label{eqn:rt}       
\eqaend
where the curvature scalar $R_h$  refers to the conformally transformed
 \lq\lq metric" $h_{\alpha\beta}$.  By choosing a measure appropriately
 and going through an analysis similar to the one performed by 
David and Distler and Kawai \cite{DDK} for non-critical string theory, 
one arrives at a quantum model.  For the special case with 24 matter 
scalars, the model simplifies considerably and in the \lq\lq conformal
 gauge" $h_{\alpha\beta} = \e^{2\rho}\eta_{\alpha\beta}$  the action 
takes the form 
\eqabegin
 S &=& S^{cl} + S^{gh}\comma \\
 S^{cl} &=& {1\over \ff} \int d^2x \left( -\del_\alpha\Phi\del^\alpha\Phi
 - 2\del_\alpha\Phi\del^\alpha\rho +4\mu^2\e^{\Phi + 2\rho} 
 + \half \del_\alpha \vec{f} \cdot \del^\alpha \vec{f} \right) \period
 \label{eqn:cl}
\eqaend
where $S^{gh}$ is the usual $b$-$c$ ghost action. This is the model which 
we solved exactly in our previous work by means of a quantum 
canonical mapping into free fields. \parsmallskip
From the equations of motion, the dilaton field $\Phi$ and the Liouville 
field $\rho$ can be expressed in terms of periodic free fields 
$\psi$ and $\chi$ as 
\eqabegin
 \Phi &=& -\chi -AB \comma\label{eqn:Phisol} \\
 \rho &=& \half ( \psi -\Phi )\comma \label{eqn:rhosol}
\eqaend
where the functions $A(\xplus)$ and $B(\xminus)$ are defined by
\eqabegin
 \delplus A(\xplus)&=& \mu \e^{\psi^{+/2}(\xplus)} \comma 
\label{eqn:Aeq}\\
 \delminus B(\xminus) &=& \mu\e^{\psi^{-/2}(\xminus)}\period 
\label{eqn:Beq}
\eqaend
( In Eq.(\ref{eqn:Phisol}) the sign of $\chi$ is reversed compared with 
 I. It is not difficult to check that this is the only change 
necessary to be consistent with the original CGHS action we adopt in 
 this article. )  The light-cone coordinates are defined as usual by 
$x^\pm = t\pm\sigma$ and $\psi^{\pm/2} (x^\pm)$ are the left- and right-
 going components of the free field $\psi(x)$. 
 We write the Fourier mode expansions of $\psi$ and 
$\chi$ as 
\eqabegin
\psi &=& \gt\left\{ \qplus 
 + \pplus(\xplus +\xminus) + i\sum_{n\ne 0} 
 \left({\alpha^+_n \over n}\e^{-in\xplus} 
+{\tilde{\alpha}^+_n \over n}\e^{-in\xminus} \right)\right\} \comma\\
\chi &=& \gt\left\{ \qminus 
 + \pminus(\xplus+\xminus) + i\sum_{n\ne 0} 
 \left({\alpha^-_n \over n}\e^{-in\xplus} 
+{\tilde{\alpha}^-_n \over n}\e^{-in\xminus}\right)\right\} \comma
\eqaend
where  $\gt$, to be often used hereafter,  is defined as 
\eqabegin
 \gt &\equiv& {\gamma \over \sqrt{4\pi}} \period
\eqaend
Then $\psi^{+/2}$ for example takes the form 
\eqabegin
\psi^{+/2} &=&\gt \left\{ {\qplus \over 2}
 + \pplus\xplus  + i\sum_{n\ne 0} {\alpha^+_n \over n}\e^{-in\xplus} 
 \right\} \period
 \eqaend
A somewhat peculiar superscript $\pm/2$ on $\psi$ is designed to 
remind us that its zero-mode part contains $\qplus/2$, \ie half 
the corresponding part in the full $\psi$.  On the other hand,  
we will need  chiral free fields with {\it full} zero-mode 
structure, which possess better conformal properties.  These will be 
 denoted with the usual superscript $\pm$.  For instance, we define 
\eqabegin
\psi^+ &\equiv& \gt \left\{ \qplus 
 + \pplus\xplus  + i\sum_{n\ne 0} {\alpha^+_n \over n}\e^{-in\xplus} 
 \right\} \period
\eqaend
%
%%%%%%
As $\psi^{\pm/2}$ each experiences a constant shift under 
$\sigma \rightarrow
 \sigma + 2\pi$, $A(\xplus)$ and $B(\xminus)$ are not periodic and 
 satisfy the boundary conditions 
\eqabegin
 A(\xplus +2\pi) &=& \alpha A(\xplus) \comma\\
 B(\xminus -2\pi) &=& {1\over \alpha} B(\xminus) \comma
\eqaend
where $\alpha$ is related to the zero mode $\pplus$ by 
\eqabegin
 \alpha &=& \e^{\gamma\sqrt{\pi}\pplus} \period
\eqaend
Solutions for $A$ and $B$ which satisfy the proper boundary conditions 
are ( suppressing the $t$ dependence )
\eqabegin
 A(\sigma) &=& \mu C(\alpha) \int_0^{2\pi} d\sigma'
    E_\alpha(\sigma-\sigma') \e^{\psi^{+/2}(\sigma')} 
\comma\label{eqn:A}\\
 B(\sigma) &=& \mu C(\alpha)\int_0^{2\pi} d\sigma''
   E_{1/\alpha}(\sigma-\sigma'') \e^{\psi^{-/2}(\sigma'')} \comma
\eqaend
where $C(\alpha) =1/\left(\alpha^{1/2}-\alpha^{-1/2}\right)$ 
 and the functions $E_\alpha(\sigma)$ and $E_{1/\alpha}(\sigma)$
 are defined by
\eqabegin
  E_\alpha(\sigma) &\equiv& 
      \exp\left(\half\epsilon(\sigma)\ln\alpha\right) \comma \qquad
  E_{1/\alpha}(\sigma) \equiv 
      \exp\left(-\half\epsilon(\sigma)\ln\alpha\right) \period
\eqaend
$\epsilon(\sigma)$ is a stair-step function with the property
$\epsilon(\sigma +2\pi) = 2 +\epsilon(\sigma) $ and  coincides with the 
usual $\epsilon$-function in the interval $[-2\pi, 2\pi]$. 
Note that we must require $\pplus$ not to vanish since otherwise 
 $C(\alpha)$ blows up.  \parsmallskip
The left-going and the right-going energy-momentum tensor $T_{\pm\pm}$
 take simple forms in terms of the free fields.  With a convenient 
normalization, they take the form
\eqabegin
  T_{\pm\pm} &=& {1\over \gt^2}\left( 
\del_\pm \chi \del_\pm \psi -\del_\pm ^2\chi
   + \half (\del_\pm \vec{f})^2 \right) \period \label{eqn:fEMT}
\eqaend
This can be diagonalized by introducing canonically normalized scalar fields
 $\phi_1$, $\phi_2$ and $\phi_f$: 
\eqabegin
\psi &=& {\gt \over \sqrt{2}}\left(\phi_1 +\phi_2 \right)
 \comma \qquad
\chi = {\gt \over \sqrt{2}}\left(\phi_1 -\phi_2 \right)
 \comma \qquad
f^i = \gt\phi_f^i \comma \\
T_{\pm\pm} &=& \half\left(\del_\pm\phi_1\right)^2 
  -Q\del_\pm^2\phi_1 - \half\left(\del_\pm\phi_2\right)^2 
  +Q\del_\pm^2\phi_1 + \sum_{i=1}^{24}
  \half\left(\del_\pm\phi_f^i\right)^2 \comma\label{eqn:pEMT}
\eqaend
where the background charge $Q$ is given by 
\eqabegin
Q &=& {\sqrt{2\pi}\over\gamma} ={1\over \sqrt{2}\gt}\period
\eqaend
Fourier mode expansion for $f^i$ is just like for $\psi$ with 
 the replacements $(\qplus,\, \pplus,\, \alplus_n,\,
\tilde{\alplus}_n)$ $\longrightarrow$ $(q^i_f,\,p^i_f,\, \alpha^i_f,\,
 \tilde{\alpha}^i_f)$.  \parsmallskip
%%%%%%%%
As was fully described in I, one can show that the mapping from the 
original fields $\{\Phi, \rho\}$ into the free fields 
$\{\psi, \chi\}$ is a quantum as well as classical canonical transformation.
 Namely, the canonical equal time commutation relations 
 \eqabegin
  \left[\Phi(\sigma, t), \Pi_\Phi( \sigma', t)\right] &=& 
  \left[\rho(\sigma, t), \Pi_\rho (\sigma', t)\right] = i\hbar
\delta(\sigma -\sigma') \nn
  \eqaend
are reproduced if we assume the commutators among the modes of $\psi$ and 
 $\chi$ to be 
\eqabegin
 \left[q^\pm, p^\mp\right] &=& i\hbar \comma \\
 \left[\alpha^\pm_m, \alpha^\mp_n\right] &=& 
 \left[\tilde{\alpha}^\pm_m, \tilde{\alpha}^\mp_n\right] = m\hbar 
 \delta_{m+n, 0} \comma \\
 \mbox{Rest} &=& 0 \period
 \eqaend
(Commutators between the modes of $f^i$ are of the usual form.)
 To establish this result quantum mechanically, it was important that 
 the non-local operators $A(\xplus)$ and $B(\xminus)$ are well-defined 
 without the need of normal-ordering due to the commutativity of the 
 modes of $\psi$. \parsmallskip
 %%%%%
 The quantized model continues to enjoy conformal invariance. 
The Fourier   modes $L_m$ and $\bar{L}_m$ of the left- and right-going 
energy-momentum   tensors in the dilaton-Liouville (dL) 
   and the matter (f) sector satisfy the usual Virasoro algebra with the 
 central charges $c^{dL} = 2$ and $c^f=24$ respectively. 
 It was shown in I  that $\chi$ and  the product 
$A(\xplus)B(\xminus)$ tranform as 
genuine dimension  zero primary fields,   while due to the presence of
 the background charge   $\psi$ transforms anomalously as 
 \eqabegin
 \left[L^{dL}_m, \psi(x)\right] &=& \e^{im\xplus}\left(\overi \delplus
 \psi + {Q \gamma \over \sqrt{2\pi}}m \right) \period 
\label{eqn:psitrans}
 \eqaend
In the subsequent sections, {\it chiral} primary fields with 
dimension 0 will play important roles.  $\chi^+$ ( not $\chi^{+/2}$ )
 is one such field. Another one is a slight modification of $A(\xplus)$ 
defined by 
\eqabegin
 \calA(\xplus) &\equiv & \e^{\gt\qplus/2}A(\xplus)
\period \label{eqn:calA}
\eqaend
The additional zero-mode factor makes this field transform as a 
genuine chiral primary field. 
 \parsmallskip
 %%%%%%%%%
 As far as the mathematical structure is concerned, our model is 
 a hybrid of critical and non-critical  bosonic string theories. 
 Hence the physical states are readily obtained by using the BRST analysis 
 developed for these theories with appropriate modifications
\cite{LZ}-\cite{MS},\cite{HKS}. 
 Let us briefly summarize the results obtained in I. ( Only the left-going
  sector will be treated explicitly. )
 \parsmallskip
 %%%
 The nilpotent BRST operator is given by
 \eqabegin
 d &=& \sum c_{-n}\left(\Ldl_n+L^f_n\right) -\half 
\sum :(m-n)c_{-m}c_{-n}b_{m+n} : \comma \label{eqn:BRSop}
\eqaend
where $sl(2)$ invariant normal-ordering for the ghosts is assumed. 
The physical ghost vacuum is defined as usual by
 $ \downvacket = c_1 \mid 0 >_{inv} $.
The operator $d$ is decomposed 
 with respect to the ghost zero mode in the form 
$d = c_0 \Ltot_0 -Mb_0 + \dhat$, where $\Ltot_0$ is the total Virasoro 
operator  including the ghosts.  It is well-known that the non-trivial 
$d$-cohomology must be in the sector satisfying $\Ltot_0\psi =0$.
By assigning the degree to the mode operators
\eqabegin
 \mbox{deg}(\alplus_n ) &=& \mbox{deg}(c_n) =1 \comma 
 \qquad
 \mbox{deg}(\alminus_n ) = \mbox{deg}(b_n) =-1 \comma \nn\\
 \mbox{deg(Rest)} &=& 0 \comma
\eqaend
the BRST operator for the relative cohomology $\dhat$ is decomposed 
 as
\eqabegin
 \dhat &=& \dhat_0 + \dhat_1 +\dhat_2\comma \\
 \dhat_0 &=& \sum_{n\ne 0} P^+(n)c_{-n}\alminus_n \comma \\
 \dhat_1 &=& \sum_{n.z.m.}:c_{-n}( \alplus_{-m}\alminus_{m+n}
     + \half (m-n)c_{-m}b_{m+n} +L^f_n) : \comma\\
 \dhat_2 &=& \sum_{n\ne 0} P^-(n)c_{-n}\alplus_n \comma
\eqaend
where $P^\pm(n)$ are given by
\eqabegin
 P^+(n) &=& \pplus + i\sqrt{2}Qn \comma \qquad 
 P^-(n) = p_- \period
\eqaend
One then studies $\hat{d}_0$- or 
$\hat{d}_2$-cohomology depending on the conditions on $P^\pm(n)$, and 
upon them all the $\hat{d}$- and $d$- cohomologies can be constructed. 
\parsmallskip
%%%%%%%%%%%
As our purpose in this article is to extract space-time geometry of states 
in which the matter fields carry finite energy in the limit of large $L$, 
we only record the relevant $d$-cohomologies, namely the ones with 
arbitrarily high matter excitations without ghosts.  Let $\psi_0$ be 
a state of the form
\eqabegin
\psi_0 &=& \hat{F} \pdownvac \comma\\
\vec{P} &=& =(\pplus, \pminus, \vec{p}_f) \comma \\
\Ltot_0\pdownvac &=& \left(\pplus\pminus + \half \vec{p}^2_f-\hbar\right)
\pdownvac =0 \comma
\eqaend
where $\hat{F}$ is an operator composed of arbitrary number of matter 
creation operators.   Such a state simultaneously belongs to $\hat{d}_0$- 
and  $\hat{d}_2$-cohomologies.  If $P^+(n)\ne 0$ for all non-zero integer 
$n$, the corresponding physical state $\psi$ satisfying $\Ltot_0\psi =0$ 
can be constructed in the form
\eqabegin
 \psi &=& \sum_{n=0}^\infty (-1)^n (T^+)^n \psi_0 \comma
\label{eqn:dhcohomp} \\
 T^+ &\equiv & \invNhat \Kplus \dhat_1 \comma \label{eqn:Tplus} \\
\Kplus &\equiv& \sum_{n\ne 0} {1\over \Pplus(n)} \alplus_{-n} b_n \comma
 \label{eqn:Kplus}\eqaend
where $\hat{N}_{dLg}$ is the level counting operator for the 
dilaton-Liouville-ghost sector.  Similarly, if $\pminus \ne 0$, the 
 expression for $\psi$ becomes
\eqabegin
 \psi &=& \sum_{n=0}^\infty (-1)^n (T^-)^n \psi_0 \comma
 \label{eqn:dhcohomm} \\
 T^- &\equiv & \invNhat \Kminus \dhat_1 \comma \label{eqn:Tminus}\\
 K^-&\equiv& {1\over p^-}\sum_{n\ne 0}\alminus_{-n}b_n \period
 \label{eqn:Kminus}
\eqaend
In the next section, we shall give more useful representations for these 
somewhat formal expressions. 
%%%%%%
%%%%%%%%%%%%%%%%%%%%%%%%%%%
\section{DDF Representations of Physical States}
%%%%%%%%%%%%%%%%%%%%%%%%%%%%
\sectionnumbering
\sindent
As we have seen, the structure of physical states of our model is 
formally extremely similar to that of bosonic string theories. 
  However, the physical interpretation of them is quite 
different. In string theory,  Virasoro levels specify the invariant
 masses of various fields, while in the present model they refer to the 
discretized energy levels of a field: The energy 
carried by a state at level $n$ is proportional to $n/L$.  As we will 
be most interested in  configurations where the matter fields 
carry finite energy in the limit of large $L$, we need to be able to 
deal with physical states at arbitrary high Virasoro levels, in marked 
contrast to the case of string theory. \parsmallskip
%%%%%%
For this purpose, the formal expressions (\ref{eqn:dhcohomp}) and 
(\ref{eqn:dhcohomm}) of physical states 
obtained through BRST analysis are not readily tractable. Fortunately, 
for states not involving ghosts, more useful expressions are available,
 the so called DDF states \cite{DDF}, developed long ago in the context of 
string theory. 
Let us briefly describe the essence of the construction in a manner
 suitable for our model. \parsmallskip
%%%%%%%%% 
Let $\phi^i_f(\xplus)$ be  canonically normalized left-going matter 
fields and $\varphi(\xplus)$ be a dimension 0 primary field with the 
following properties:
\eqabegin
&& (i)\quad \e^{im\varphi} \mbox{\ is periodic for $m\in {\bf Z}$} 
\comma\nn\\
&& (ii) \quad \mbox{modes of $\varphi$ all commute with themselves 
 and with the matter fields $\phi^i_f$ } \comma \nn\\
&& (iii) \quad \int_0^{2\pi}{d\xplus \over 2\pi} \delplus \varphi(\xplus)
 = 1 \period \nn
\eqaend
Then the set of operators $B^i_m$ defined by 
\eqabegin
 B^i_m &\equiv& \int_0^{2\pi}{d\xplus \over 2\pi} \e^{im\varphi}
\delplus\phi^i_f 
\eqaend
are BRST invariant and satisfy the oscillator commutation relations 
\eqabegin
\left[ B^i_m, B^j_n\right] &=& \hbar m\delta_{m+n,0} \period
\eqaend
BRST invariance is trivial since the integrand of $B^i_m$ is a periodic
dimension 1 primary field and the commutation relations also follow
 easily using the properties $(i)\sim (iii)$ listed above. In particular,
 the periodicity requirement is crucial in order to perform the 
 integration by parts during the course of the calculation. 
\parsmallskip
%%%%
For our model,  the simplest canditate for $\varphi$ ( when $\pminus 
 \ne 0$ ) is $\varphi(\xplus)=\chi^+(\xplus)/(\gt\pminus)$ and the 
 corresponding $B^i_m$, which we denote by $A^i_m$, is given by
\eqabegin
 A^i_m &=& \int_0^{2\pi}{d\xplus \over 2\pi} \e^{im
\chi^+(\xplus)/(\gt\pminus)}\delplus\phi^i_f \period
\eqaend
The factor of $1/\pminus$ is necessary to assure the periodicity. 
Physical states can be built up using these oscillators 
as
$$ \sum C_{i_1i_2\cdots i_k}^{n_1n_2 \cdots n_k}A^{i_1}_{-n_1}
A^{i_2}_{-n_2}\cdots A^{i_k}_{-n_k}\pdownvac \comma $$
where $\pdownvac$ is the zero-mode vacuum satisfying  the condition
$\pplus\pminus +(1/2)\vec{p}_f^2 -\hbar=0$. A characteristic feature of 
this type of physical states is that they are composed solely of 
the oscillators $\alminus_{-n}$ and $\alpha^i_{-m}$ and do not contain 
$\alplus_{-n}$'s.  In the BRST formalism, this property is precisely 
the one enjoyed by the states of the form in (\ref{eqn:dhcohomm}), 
namely, 
$$ \sum_{n=0}^\infty \left(-T^-\right)^n \psi_0 \comma $$
where $\psi_0$ is a state representing $\hat{d}_2$-cohomology with 
matter excitations only. Indeed one can show that they are identical.
The precise identification is 
\eqabegin
 \vec{a_1} \cdot \vecA_{-k_1} \vec{a_2} \cdot \vecA_{-k_2} \cdot \cdot 
\cdot
 &\vec{a_q}& \cdot \vecA_{-k_q} {\mid \vec{P'} >_\downarrow} \nn \\
 &=&\sum_{n=0}^{\infty}(-T^-)^n  \vec{a_1} \cdot \vecalf_{-k_1} 
\vec{a_2} 
 \cdot \vecalf_{-k_2} \cdot \cdot \cdot \vec{a_q} \cdot \vecalf_{-k_q} 
\pdownvac \comma \label{eqn:DDF}
\eqaend
where
\eqabegin
&& p'_+p'_- + \frac12 {\vec{p}'_f}{}^2 -\hbar = 0 \comma \qquad
 \pplus \pminus + \frac12 \vecpf{}^2 + \hat{N}_f -\hbar = 0 \comma \nn \\
&& p'_+ +{\hbar \over p'_-}(k_1+\cdot\cdot\cdot+k_q)=\pplus \comma 
\qquad  p'_-= \pminus \comma \qquad \vec{p}'_f= \vecpf \comma \nn 
\eqaend
and $\vec{a}_i$'s are arbitrary constant vectors. 
( The difference in the zero-mode sector is simply due to the fact that 
$A^i_{-m}$ contains the factor $\exp(-im\qminus/\pminus)$ which 
shifts $\pplus$ by the amount $-m\hbar/\pminus$, while $T^-$ 
 does not contain such zero-mode operator. ) \parsmallskip
In fact the proof of this formula is already implicit in our 
previous work, namely in the proof of Eq.(4.24)( with techinical 
details  in Appendix B ) in I, with the replacement of $T^+$ by $T^-$ and 
some associated changes. We now make its relevance more explicit. 
\parsmallskip
Let $\psi_0$ be a state made up solely of 
matter oscillators as stated above. Hence it is of degree 0 and is 
annihiliated by both $\hat{d}_0$  and $\hat{d}_2$. The problem is  
to construct a representative $\psi$ of $\hat{d}$-cohomology, which 
contains $\psi_0$. Since we are interested in $\psi$ composed only of 
$\alminus_{-n}$ and $\alpha^i_{-m}$, the degree of  $\psi$ must be 
 non-positive and we can expand it as 
\eqabegin
 \psi &=& \sum_{n\ge0} \psi_{-n} \comma
\eqaend
where ${\rm deg}(\psi_{-n}) = -n$. Then the $\hat{d}$-closedness 
condition for $\psi$ reads
\eqabegin
 \hat{d}\psi &=& \left( \hat{d}_0 + \hat{d}_1 + \hat{d}_2\right)
 \sum_{n\ge0} \psi_{-n} \nn\\
 &=& \sum_{n} \left( \hat{d}_0\psi_{-n} + \hat{d}_1\psi_{-(n+1)}
 + \hat{d}_2 \psi_{-(n+2)}\right) \\
 &=& 0 \comma \nn
\eqaend
which leads to the recursion relations
\eqabegin
 \hat{d}_1 \psi_0 + \hat{d}_2 \psi_{-1} &=& 0 \comma \\
 \hat{d}_0\psi_{-n} + \hat{d}_1\psi_{-(n+1)}
 + \hat{d}_2 \psi_{-(n+2)} &=& 0 \qquad (\mbox{for }\ n\ge 0)\period
\eqaend
By an argument parallel to that given in Appendix B of I, 
 provided $\pminus \ne 0$, one can show recursively that 
 $K^-\psi_{-n}=0$ and $\hat{d}_0
 \psi_{-n} =0$.  This latter statement means that indeed
 $\psi$ does not contain $\alplus_{-n}$ oscillators. Then the 
recursion relation simplifies to 
\eqabegin
  \hat{d}_2 \psi_{-(n+1)} &=& -\hat{d}_1\psi_{-n} 
\eqaend
and the rest of the argument in I amounted to showing that, within the
 space of states without the excitations of ghosts and $\alplus_{-n}$
 oscillators,  $\hat{d}_2$ has the inverse 
 given by $\hat{N}^{-1}_{dLg}K^-$ and hence the recursion 
relation is {\it uniquely} solved starting from $\psi_0$. Therefore to 
prove the validity of Eq.(\ref{eqn:DDF}), all one has to do is to check 
that the degree zero part of the both sides are identical, but this is 
trivial. 
%%%%%%%%%%%%%%%%%%%%%%%%%%%
\parsmallskip
From the argument just presented, it is clear that when $P^+(n) \ne 0$ 
physical states (\ref{eqn:dhcohomp}) constructed in terms of 
$T^+$ must also have DDF type representation. To find it,  one must 
look for a candidate for  the periodic dimension 0 primary $\varphi$, 
which consists only of 
the modes of $\psi^+(\xplus)$. $\psi^+(\xplus)$ itself, however, is not 
appropriate since it does not transform as a primary field due to the 
presence of the background charge ( cf. Eq.(\ref{eqn:psitrans}) ). 
 The correct choice of $\varphi$  turns out to be 
\eqabegin
\varphi &=& {1\over \gt \pplus}\etap \comma\\
\etap &\equiv&   \ln \left( \calA(\xplus)/\mu \right) \comma
\label{eqn:etap}
 \eqaend
where $\calA(\xplus)$ is a genuine primary field of dimension 0 
 defined in Eq.(\ref{eqn:calA}). Notice that  $\pplus$ must 
 not  vanish for this construction, but this condition is already 
 needed in defining $A(\xplus)$ and $B(\xminus)$. 
A useful explicit form of $\etap$ is derived in Appendix A, together 
with its conjugate denoted by  $\zeta^+$. 
The fields $\eta^+$ and 
 $\zeta^+$ are intimately related to the ones employed in 
\cite{VV1,VV2}. \parsmallskip
 Thus physical states can be 
built up by the BRST invariant oscillators 
\eqabegin
 \At^i_{-n} &\equiv & \e^{-i(n /\gt \pplus)
\ln(\gt \pplus)}
\int_0^{2\pi}{d\yplus \over 2\pi} \e^{-in
 \etap/\gt\pplus} \delplus\phi_f(\yplus) \period \label{eqn:Atil}
 \eqaend
The extra phase factor in front, which commutes with the BRST operator, 
 is added to remove the corresponding phase in the integrand so that 
the physical states built with these oscillators agree with the 
ones constructed with $T^+$ operators. 
In the next section, we shall make use of this type of oscillators to 
construct interesting physical states. 
%%%%%%%%%%%%%%%%%%%%%%%%%%%%%
%%%%%%%%%%%%%%%%%%%%%%%%%%
\section{ Extraction of Space-Time Geometry}
%%%%%%%%%%%%%%%%%%%%%%%%%%%%
\sectionnumbering
\sindent 
As was already pointed out in the introduction, physical meaning of an 
abstract state can only be extracted by looking at its response to the 
action of appropriate operators of physical significance.  
In quantum gravity, each physical 
state corresponds to a possible choice of the universe and
 all the events which \lq\lq take place " in that universe 
must already be encoded in a chosen state.  This  means 
 that there is no meaning to  a \lq\lq transition " between different 
physical states and consequently we will be interested only in the 
average values of suitable operators in a particular physical state. 
\parsmallskip
%%%%%%%%%%
\subsection{Choice of Probing Operators }
%%%%%%%%%%%%%%%%%
\sindent
The first question then is which operators are suitable for probing the 
content of a physical state.  Preferably we wish to use an appropriate 
set of BRST invariant operators since their expectation values are 
independent 
of the choice of the representative of the physical state.  They are 
essentially the integrals of dimension 1 vertex operators familiar 
in string theory.  As they are manifestly coordinate-independent, their 
expectation values are simply a set of numbers. In string theory, 
these set of numbers have immediate physical significance; they are 
functions of the momenta of the particles which propagate in the target 
space.  In quantum gravity context, however, they are very hard to interpret.
 One might try to draw an analogy to the description of a 
charge distribution in terms of a set of integrals, namely the 
 multipole moments.  A crucial distinction is that in that
case a definite physical picture is already attached to the 
functions forming the basis of the expansion and with the knowledge of 
 the values of the moments we can immediately reconstruct the physical 
distribution.  Here we do not have such an underlying expansion.  Thus 
 although one cannot deny a possibility that in the future, with enough 
 experience and ingenuity, we  may be able to understand 
physics  directly from an infinite set of gauge invariant numbers, but at 
present it is obviously not productive to pursue such a route. 
\parsmallskip
We shall then try to deal with operators of more direct physical 
significance,  such as the stress-energy tensor of the matter fields,
 the metric,  and the curvature.  As for the latter two entities, 
there are some ambiguities:  First of all, it is not clear which of the two 
conformally related quantities, $g_{\alpha\beta}$ and $h_{\alpha\beta}$, 
 should be regarded as the metric. This question of the choice of
 \lq\lq conformal-frame " often occurs in dilaton gravity and the principle
  of general coordinate invariance alone cannot dictate the correct choice. 
We shall therefore keep both possibilities open.  
 Clasically, from the 
definitions (\ref{eqn:Phih}), (\ref{eqn:omega}) and the canonical 
transformation (\ref{eqn:Phisol}),(\ref{eqn:rhosol}), the metric and the 
curvature in these two schemes can be expressed in terms of the free 
fields ( in the original coordinate $\xi^\mu$ ) as
\eqabegin
g_{\alpha\beta} &=& \Phi^{-1}\e^\psi \eta_{\alpha\beta} \comma
\qquad 
g^{\alpha\beta} = \Phi\e^{-\psi}\eta^{\alpha\beta} \comma \\
R^g_{\alpha\beta} &=& \half \Box \ln \Phi \eta_{\alpha\beta}
 \comma \qquad 
R^g = \e^{-\psi}\Phi\Box\ln\Phi \comma \\
h_{\alpha\beta} &=& \e^{\psi -\Phi}\eta_{\alpha\beta} \comma 
\qquad 
h^{\alpha\beta} = \e^{\Phi -\psi }\eta^{\alpha\beta}\comma \\
R^h_{\alpha\beta} &=& -2\lambda^2\e^\psi \eta_{\alpha\beta} \comma 
\qquad 
R^h = -4\lambda^2\e^\Phi \period 
\eqaend
Due to the composite nature and the presence of the complicated 
expression 
$\Phi =-\chi -AB$, it is not an easy task to give proper quantum 
definitions 
for these operators.  In this article, we shall treat two 
 of the relatively simple ones, namely $g^{\alpha\beta}$ and 
$R^h_{\alpha
 \beta}$. 
Recalling that all the modes of $\psi$ commute with each other, 
$R^h_{\alpha\beta}$ and $-AB \e^{-\psi}$ part of $g^{\alpha\beta}$ 
is already well-defined. On the other hand, the remaining part of the 
latter 
operator, namely $\chi\e^{-\psi}$, needs regularization. For this 
purpose, let us decompose $\chi$ and $\psi$ into the zero-mode, 
 the annihiliation, and the creation parts:
\eqabegin
 \chi &=& \chi_0 +\chia + \chi_c \comma\\
\psi &=& \psi_0 + \psi_a +\psi_c = \psi_0 + \tilde{\psi} \comma
\eqaend
where $\tilde{\psi}$ denotes the non-zero-mode part of $\psi$. 
We can then define the operator $\nchiepsi$ by the 
normal-ordering:
\eqabegin
 \nchiepsi &=& \chi_0 \e^{-\psi_0}\e^{-\psit} + \chi_c\e^{-\psi}
 + \e^{-\psi} \chia \period
\eqaend
For the first term, we have written out the zero-mode part explicitly.
 It is easy to check that $\chi_0$ and $\psi_0$ commute with each
 other and hence $\nchiepsi$ defined above is properly hermitian. 
Therefore, we can actually write 
\eqabegin
 \nchiepsi &=& \chi \e^{-\psi} -\left[\chia, \e^{-\psi}\right]\period
\eqaend
Conformal property of $\nchiepsi$ is easily worked out to be
\eqabegin
 \left[L_n, \nchiepsi\right] &=& \hbar \e^{in\xplus}
 \left( \overi \delplus -n \right) \nchiepsi \nn\\
 && + \hbar^2 \gt^2 n \e^{in\xplus} \e^{-\psi} \period 
\label{eqn:Lnchiepsi} 
\eqaend
This shows that the conformal tranformation property of 
regularized $g^{\alpha\beta}$ is slightly modified by a higher order 
 contribution and it is no longer a conformal 
primary. However, as we shall fix the gauge completely, this will 
not cause any problems. Together with the matter part of the 
energy-momentum tensor, these  operators will give interesting
 physical information. 
\parsmallskip
%%%%%%%%%%%%%%%%%%%%%%%%%%%%%%%%%%
%%%%%%%%%%%%%%%%%%%%%%%%%%%%%%%%%%%%%
\subsection{Choice of States }
%%%%%%%%%%%%%%%%%%%%%%%%%%%%%%%%%%
\sindent
Clearly the operators we have chosen to work with are  gauge 
dependent and hence their expectation values inevitably depend on the 
choice of the representative of the physical state, which amounts to a 
choice of gauge within the conformal gauge.  If we denote by 
$\ket{\Psi_0}$ a special representative of a 
 non-trivial cohomology class satisfying $L^{tot}_0 \ket{\Psi_0}=
\bar{L}^{tot}_0 \ket{\Psi_0} =0 $, 
 any other representative $\ket{\Psi}$ of the same class is 
expressed  as 
\eqabegin
 \ket{\Psi} &=& \ket{\Psi_0} + \ket{\Lambda} \comma \\
 \ket{\Lambda} &=& d \ket{\Psig}  + \dbar \ket{\Psibarg} \comma
\eqaend
where $\ket{\Psig}$ and $\ket{\Psibarg}$ are arbitrary states with 
left- and right- ghost number $-1$ respectively.
The question is how we should choose $\ket{\Psi_0}$, $\ket{\Psig}$ 
 and $\ket{\Psibarg}$ 
so that the average value $\bra{\Psi}{\cal O}(x)
\ket{\Psi}$ for the operator indicated above exhibits 
an interesting physically interpretable behavior. A hint is provided
by a simple fact about the coordinate dependence 
 of a matrix element $\bra{a} {\cal O}(x)\ket{b}$,  where the states 
 $\ket{a}$, $\ket{b}$ and the operator ${\cal O}(x)$ carry definite 
global left-right dimensions \ie 
\eqabegin
  L^{tot}_0 \ket{a} &=& \hbar \Delta_a \ket{a}\comma  \qquad 
 \bar{L}^{tot}_0 \ket{a}=\hbar \bar{\Delta}_a \ket{a} \\
  L^{tot}_0 \ket{b} &=& \hbar \Delta_b \ket{b}\comma  \qquad 
 \bar{L}^{tot}_0 \ket{b}=\hbar \bar{\Delta}_b \ket{b} \\
 \left[L^{tot}_0, {\cal O}(x)\right] &=& {\hbar \over i} 
  \delplus {\cal O}(x)\comma \qquad 
\left[\bar{L}^{tot}_0, {\cal O}(x)\right] = {\hbar \over i} 
  \delminus {\cal O}(x)  \period
\eqaend
By evaluating the matrix element 
$\bra{a}\left[L^{tot}_0, {\cal O}(x)\right]\ket{b}$ 
and a similar one with $\bar{L}^{tot}_0$, we easily deduce
\eqabegin
 \bra{a} {\cal O}(x) \ket{b} &=& const.\ 
 \e^{i\left(\Delta_a -\Delta_b\right)
 \xplus} \cdot \e^{i\left(\bar{\Delta}_a -\bar{\Delta}_b\right)
 \xminus} \comma
\eqaend
which expresses nothing but the conservation of energy and momentum. 
Let us apply this to the case of interest, namely to the expectation 
value 
\eqabegin
 \bra{\Psi}{\cal O}(x)\ket{\Psi} 
&=& \bra{\Psi_0}{\cal O}(x)\ket{\Psi_0} \nn\\
& & \quad + \bra{\Psi_0}{\cal O}(x)\ket{\Lambda} 
+ \bra{\Lambda}{\cal O}(x) \ket{\Psi_0}  \label{eqn:meanval} \\
& & \quad \quad +  \bra{\Lambda}{\cal O}(x) \ket{\Lambda}
\period \nn
\eqaend
Then we immediately learn the following:
First,  $\bra{\Psi_0}{\cal O}(x)\ket{\Psi_0}$ part can only be 
 a constant. Second, the remaining part can produce non-trivial 
coordinate dependence if $\ket{\Lambda}$ carries non-vanishing weights. 
In particular, by arranging $\ket{\Lambda}$ to be a suitable 
superposition of states with various weights, it should be possible to 
generate a wide variety of coordinate dependence. \parsmallskip
%%%%%
To further narrow down the appropriate choice of $\ket{\Psi_0}$ and 
$\ket{\Lambda}$, let us note that the second line of (\ref{eqn:meanval})
contains the information of the non-trivial part of the physical state 
while the last line depends only on the gauge part $\ket{\Lambda}$. 
Thus it is natural to try to choose $\ket{\Lambda}$ such that the 
interesting  coordinate dependence comes predominantly from the cross
 terms in the second line. 
As for $\ket{\Psi_0}$, various choices can 
be possible. It would however be most interesting if we can produce a 
shock-wave like configuration  for the matter energy-momentum tensor 
since then we should  see the formation of a black hole in the mean 
value of the metric. 
 We expect that such a macroscopic configuration can be constructed 
  if $\ket{\Psi_0}$ is taken to be a suitable coherent state.
\parsmallskip
%%%%%%%%%%%%%%%%%%%%%%%%%
Guided by the reasoning  above, we have chosen to work with the 
following class of states.  First, $\Psizeroket$ is taken to be 
a coherent state built up with the BRST invariant oscillators 
$\Atil_{-n}$ introduced in the previous section:
\eqabegin
\Psizeroket & \equiv & e^G \psmearvac \comma  \\
G &\equiv & {1\over \hbar}\sum_{n\ge 1}{ \tilde{\nu}_n\over n}
\Atil_{-n} \comma \\
\tilde{\nu}_n &=& \nu_n \e^{in x^+_0 }
\qquad (\nu_n,\, x^+_0\,:\, \mbox{real constants}) \comma \\
\psmearvac & \equiv & \e^{-ic\pplus /\hbar \gamma}
 \gamma^2 \int_{-\infty}^{\infty} dp^+ 
      \int_{- \infty}^{\infty} dp^- \, \weight 
      \mid p^+,p^-,\vec{p}_f >_\downarrow \nn \\
& & \times \delta (p^- - \frac1{p^+}(\hbar - \half p_f^2))\comma 
 \label{eqn:psmearvac} \\
 c &=& \mbox{ a real constant } \comma \\
p_f^2 &\equiv & \vec{p}_f \cdot \vec{p}_f \period
\eqaend
Some explanations are in order:\ 
 For simplicity, we consider a coherent state in which only one kind
 of  matter field is excited. Thus we omit the superscript $i$ on 
 $\tilde{A}_{-n}$.  
 $\psmearvac$ is a zero-mode vacuum smeared with a real weight 
 $\weight$ and it clearly satisfies  $L^{tot}_0 \psmearvac =
  \break \Lbartot 
\psmearvac=0$. This smearing is necessary to make the mean value of 
 the operator $\qminus$ well-defined, which will appear in 
$<\, g^{\alpha\beta}\, >$. 
An appropriate choice of $\weight$  will be given in Sect.5. 
 The phase factor in front is a BRST invariant and will 
be seen to produce a coordinate-independent contribution in the 
mean value $< g^{\alpha\beta}> $ and the constant $c$ will 
 be adjusted to cancel certain unwanted terms.
The phase factor in the definition of $\tilde{\nu}_n$ will produce a 
 shift $\xplus \rightarrow \xplus -x^+_0$ in certain terms  
  and  will eventually specify where a matter shock wave will 
traverse. Next,  the reason for employing  $\Atil_{-n}$,
 rather than the apparently simpler $A_{-n}$, is two-fold. First, 
 $A_{-n}$ consists of $\alminus_m$ oscillators, which 
 have non-vanishing commutators with $\e^{\pm \psi}$ contained 
 in $g^{\alpha\beta}$ and $R^h_{\alpha\beta}$. Consequently,
 when $A_{-n}$ is exponentiated to make a coherent state, calculations 
 will become practically intractable. The second and more strategic 
reason is that with this choice the energy balance between the matter
 sector and the dilaton-Liouville sector will be predominantly between 
 $(1/2)(\delplus \vec{f})^2$ and $ \delplus^2 \chi$, which is 
 precisely the situation that prevails in the gauge $\psi=0$ 
often used  in  (semi-)classical discussions 
\footnote{ In terms of the original variables, this means $\phi
 =\rho_g$, where $\phi$ and $\rho_g$ are, respectively, the dilaton
 and the Liouville mode of the metric $g_{\alpha\beta}$}. 
 These remarks
 will be substantiated when we display the result of our calculation. 
In any case, $\Psizeroket$ so constructed clearly satisfies 
an equation characteristic of a coherent state, namely
\eqabegin
\Atil_m \Psizeroket &=& \tilde{\nu}_m \Psizeroket 
\qquad ( m\ge 1 )\period
\eqaend
We also wish to call the attention of the reader that $\ket{\Psizero}$ 
contains, apart from the zero-modes, only the left-going oscillators. 
This is due to our intention to produce a left-going matter shock
 wave as treated in CGHS. \parsmallskip
%%%%%%%%%%%%%%%%%%%%%%%%%%%%%
Let us now come to the choice of the gauge part $\ket{\Lambda}$.  
We have chosen it to realize all the required features discussed above 
in the simplest possible form. It is written in the form 
\eqabegin
\ket{\Lambda} &=& {1\over \kappa} \left(d\, b_{-M}
 + \dbar\, \bbar_{-M}\right)\Omegaket \comma
\eqaend
where $b_{-M}$ and $\bbar_{-M}$ are, respectively, the left- and 
right-going anti-ghost oscillator at level $M$ and $\kappa$ is a 
constant carrying the dimension of $\hbar^2$. As long as it is finite, 
the choice of $M$ does not make any qualitative difference. 
 $\Omegaket$ is 
 chosen to be a superposition of zero-mode states of the form
\eqabegin
\Omegaket & \equiv & \sum_{k= - \infty }^{\infty} \omega_k 
\sum_{l= \pm 1 ,0}
           \mid \tilde{P}(k,l)>_{\downarrow} \comma \label{eqn:OMket} \\
\mid \tilde{P}(k,l)>  & \equiv & \, e^{-ic p^+/\hbar \gamma}
  \gamma^2 \int_{-\infty}^{\infty} dp^+ 
  \int_{- \infty}^{\infty} dp^- \, \weight \mid p^+,p^-(k,l),\vec{p}_f 
  >_{\downarrow} \nn \\
 & & \times  \delta (p^- - \frac1{p^+}(\hbar - \half p_f^2)) 
\comma  \label{eqn:Pkl} \\
      p^- (k,l) & \equiv & p^- - \frac{\hbar}{p^+} k  - 
       i \hbar \gt\, l \period  
\eqaend
%%%%%%%%%%%%%%%%%%%
$\omega_k$ are set of real coefficients. 
Notice that we sum over states with shifted $\pminus$ zero-modes. 
This is necessary for the following reasons:  First since 
$\Atil_{-n}$ contains 
 a factor $\exp(-in\qplus/\pplus)$ which shifts  $\pminus$ by the amount 
 $-n\hbar/\pplus$, the shift of the form $-\hbar k/\pplus$ is needed 
in  order to yield non-trivial overlap with $\Psizeroket$. Likewise, 
 the second shift  $-i\gt\hbar \ell $ for $\ell=\pm 1, 0$ 
 is required to produce non-vanishing results when we deal with the 
 operators $g^{\alpha\beta}$ and $R^h_{\alpha\beta}$, which contain
 $\e^{\pm \psi}$ and hence carry imaginary $\pminus$. 
 It is not  difficult to see that our choice of $\ket{\Lambda}$ has the 
 desired property that it is a superposition of states with various 
Virasoro  weights and at the same time the pure gauge part of 
the mean value is rather insensitive to the matter content. 
\parsmallskip
%%%%%%%%%%%%%%%%%%%%%%%%%%%%%%%%%%%%%%%%%%%%%
%%%%%%%%%%%%%%%%%%%%%%%%%%%%%%%%%%%%%%%%%%%%%
\subsection{Specification of Inner Product }
%%%%%%%%%%%%%%%%%%%%%%%%%%%%%%%%%%%%%%%
\sindent
Before we start our calculation of the expectation values, we must settle
 one more important issue, namely the choice of the inner product. 
In  our previous work, we have argued that the correct hermiticity 
assignments for the Fourier modes of the free fields should be the 
usual one, namely 
\eqabegin
 \alpha_n^\dagger &=& \alpha_{-n} \qquad (n \ne 0) \comma \\
 p^\dagger &=& p \comma \qquad q^\dagger = q 
\eqaend
for each field. As we emphasized there, this requirement does not yet 
fix the inner product completely. Some time ago, it was pointed out
 in \cite{AF1,AF2} that for hermitian operators with {\it continuous 
spectra} their eigenvalues need not be real and in fact 
there are two distinct classes of choices for the inner product 
which are compatible with their hermiticity. Their analysis applies to 
 our zero-mode sector and especially we must take due caution for 
 the dilaton-Liouville zero-modes $p^\pm$. If we denote them generically
 by $p$, their argument shows that one can either choose $p$ to take 
values along a \lq\lq real-like" path characterized by ${\rm Re}\, p
 > {\rm Im}\, p$ at infinity or along an \lq\lq imaginary-like" path 
defined by ${\rm Re}\, p  < {\rm Im}\, p$ at infinity. The former is a 
generalization of 
 the usual choice along the real axis and the latter actually signifies 
 the presence of indefinite metric structure of the Hilbert space in 
 question. This latter scheme was later shown \cite{AF2} to 
 correspond to the \lq\lq conformal rotation" 
 in four dimensional Euclidean quantum gravity proposed by Gibbons, 
Hawking and Perry \cite{GHP}. \parsmallskip
%%%%%%%%%%%%%%%%
In the present case, we are dealing with a Minkowski theory with an 
indefinite metric structure and occurence of complex $ p^- $ zero-modes.
( Recall the sign of the kinetic term for $ \phi_2 $ in (\ref{eqn:pEMT}) and 
  the imaginary shift present in $ p^-(k,l) $ ). 
Hence it is not obvious which of the two schemes mentioned above should 
be taken. As we describe below, the correct choice is  dictated
 by the requirement of reality of the mean 
values of various operators.    In the general formulation 
 of \cite{AF1} which allows  complex values for $p$ ( for both 
 of the two schemes ), the expectation value of an operator
 $\calO$ must be defined as $<p^* | \calO | p>$, where $p^*$ 
 is the complex conjugate of $p$. This means that even when $\calO$ is 
 hermitian its mean value need not be real.  Indeed 
 we have 
\eqabegin
\left( < \Psi(P^*) \mid \calO \mid \Psi(P) > \right)^* &=& 
            < \Psi(P) \mid \calO \mid \Psi(P^*) >  \nn \\
      &\neq&  < \Psi(P^*) \mid \calO \mid \Psi(P) >   \comma
\eqaend
where we have displayed the $ P $ dependence of $ \mid \Psi > $ 
 explicitly. An obvious way of making it real is to require 
\eqabegin
 \mid \Psi(P^*) > &=& \mid \Psi(P) > \period
\eqaend
It turns out that this condition is satisfied 
provided that we $(i)$ take $ p^+ $ and $ p^- $ to be real, 
$(ii)$ sum up the imaginary shift of $ p^- $ symmetrically with respect 
to $ l $ as in (\ref{eqn:OMket}), and $(iii)$ perform the smearing over $\pplus$ 
with $\weight$. ( The last of these procedures, 
which is already necessary to make the mean value of $\qminus$ in 
$<\, g^{\alpha\beta}\, >$ well-defined, can be seen to effect 
cancellation of a number of imaginary contributions which otherwise 
 remain.)
 If this prescription is not followed, 
one can check that  $ < g^{\alpha \beta } > $ and $ < R^h_{+-} > $ 
become complex due to the presence of $ p^+ $ or $ p^- $ in them.  
Thus the conclusion is rather 
simple : we should take the usual scheme 
\ie quantization along a ``real-like'' path.\footnote{The imaginary 
shift contained in $ p^-(k,l) $ is consistent with this choice since 
they occur only in the finite domain in the complex plane.}
We will explicitly show that all the mean values will be 
real with such an inner product. \parsmallskip
%%%%
Finally, a word should be added that in the ghost sector inclusion of 
the usual ghost zero-modes $ c_0 \bar{c}_0 $ will be implicitly 
assumed throughout.
%%%%%%%%%%%%%%%%%%%%%%%%%%%%%%%%%%%%%%%%%
%%%%%%%%%%%%%%%%%%%%%%%%%%%%%%%%%%%%%%%%%
\subsection{Calculation of Mean Values }
%%%%%%%%%%%%%%%%%%%%%%%%%%%%%%%%%%%%%%%%%%%%
\sindent
We are now ready to perform the computations of the mean values. 
 Most of the calculations are tedious but straightforward. 
Thus, we shall  sketch how we organize the calculation, 
 explain some of the non-trivial manipulations,  and then jump 
to the results.  \parsmallskip
%%%%%%%%%%%%%%%%%%%%%%%%%%55
Up to a certain point, we only need to assume that the operator $\ox$
 is hermitian and does not contain ghosts. Remembering that the 
 inner product already contains $c_0\bar{c}_0$, 
$\ket{\Psi}$ and  $\bra{\Psi}$ can be simplified to 
\eqabegin
 \ket{\Psi} &=& \ket{\Psi_0} + {\hbar\over \kappa}\calLm \Omegaket
 \comma \\
 \bra{\Psi} &=& \bra{\Psi_0} + {\hbar\over \kappa}\Omegabra
 \calLp \comma \\
\calLm &=& \Ltot_{-M} + \Lbartot_{-M} \comma \\
\calLp &=& \Ltot_{M} + \Lbartot_{M} \period
\eqaend
Note that both $\ket{\Psizero}$ and 
$\Omegaket$ are annihiliated by $\Ltot_M$ and $\Lbartot_M$ and  hence
 $\calLp \Omegaket$ $ = \Omegabra\calLm =0$  holds. 
After a simple calculation, we can then organize the mean value of 
an operator $\calO$ in the following way:
\eqabegin
 \bra{\Psi}\calO\ket{\Psi} &=& \calM_0 + \calM_1
 + \calM_2 \comma \\
\calM_0 &=& \bra{\Psi_0}\calO\ket{\Psi_0} \comma\\
\calM_1 &=& \calM_{1+} + \calM_{1-} \comma\\
\calM_{1+} &=& {\hbar\over \kappa} \Omegabra
 \left[ \calLp, \calO \right] \ket{\Psi_0}  \comma\\
\calM_{1-} &=& {\hbar\over \kappa} \bra{\Psi_0} \left[\calO,
 \calLm \right] \Omegaket = \calM_{1+}^\ast  \comma\\
\calM_2 &=& \calM_{20} + \calM_{21} \comma\\
\calM_{20} &=& {\hbar^2 \over \kappa^2}\Omegabra
  \calO \left[\calLp, \calLm\right] \Omegaket \comma\\
\calM_{21} &=& {\hbar^2 \over \kappa^2}
\Omegabra \left[ \left[ \calLp, \calO\right], 
\calLm\right] \Omegaket \period
\eqaend
%%%%%%%%%%%%%%%%%%%%%%%%
The most important and somewhat non-trivial part of the calculation is 
that of $\calM_{1+}$ for $\calO = \nchiepsi$ and $ T^f$.  Below let us 
give some details of this calculation.  \parsmallskip
%%%%
First consider the case $\calO = \nchiepsi$.  Since we have already 
 worked out the conformal property of this operator in 
(\ref{eqn:Lnchiepsi}), the commutator in $\calM_{1+}$ is easily 
obtained to be 
\eqabegin
 \left[\calLp, \nchiepsi\right] &=& 
 \hbar \Dhat(M,x)\nchiepsi + \hbar^2 f(M,x) \e^{-\psi} \comma
\eqaend
where the diffential operator $\Dhat(M,x)$ and the function $f(M,x)$
 are defined by 
\eqabegin
\Dhat(M,x) &\equiv& \e^{iM\xplus}\left(\overi \delplus -M\right)
 + \e^{iM\xminus}\left(\overi \delminus -M\right) \comma \\
f(M,x) &\equiv& \gt^2M\left(\e^{iM\xplus} + 
\e^{iM\xminus} \right) \period
\eqaend
Thus our calculation reduces to that of $\Omegabra \nchiepsi 
\ket{\Psizero}$ and $\Omegabra \e^{-\psi} \ket{\Psizero}$.  Calculation
 of the latter is rather trivial since 
 the operator $\e^{-\psi}$ contains $\alplus_n$ oscillators 
 only and they act trivially on the coherent state $\ket{\Psizero}$. 
 So we shall concentrate on the former.  Recalling the form  
 of the coherent state and using the fact that 
$\Omegabra$ consists only of zero-modes, we easily get 
\eqabegin
 \Omegabra\nchiepsi \ket{\Psizero} &=&
 \Omegabra \chi_0\e^{-\psi_0}\ket{\tilde{P}} 
 + \Omegabra\e^{-\psi_0}\chia \e^G\ket{\tilde{P}} \period
\eqaend
There is no problem evaluating the first term on the right hand 
 side, but for the second we need to develop a formula for moving 
$\chia$ through $\e^G$ to the right. 
By a standard formula, 
\eqabegin
 \chia \e^G &=& 
 \e^G\left\{ \chia -\left[G,\chia\right]
 + \half\left[G, \left[G,\chia\right]\right]+\cdots \right\}\period
\eqaend
The series actually terminates after the double commutator for the 
following reason:  $\left[G,\chia\right]$ no longer contains 
$\alminus_m$'s and also 
 it is linear in $\alf_n$. Thus the double commutator can only
 contain $\alplus_n$'s and hence the rest of the series vanishes. 
\parsmallskip
%%%%%%
Recalling $G =\sum_{n\ge 1}(\tilde{\nu}_n/\hbar n)\At_{-n}$, we first 
need to compute $\left[\alminus_m, \At_{-n}\right]$, where $\At_{-n}$ 
is given in (\ref{eqn:Atil}). 
 Since $\alminus_m$ has a non-vanishing commutator
 with $\alplus_{-m}$ and the result is a c-number, we only need 
 to compute $\left[\alminus_m, \etap\right]$.  From the expression 
 of $\etap$ obtained in  Appendix A , we get 
\eqabegin
 \left[\alminus_m, \etap\right] 
 &=& {1\over \calA} {\mu \over \gt} \e^{\psi_0^+}\sum_k {1\over 
\Pplus(k)}   \e^{ik\yplus} \left[\alminus_m, C_{-k}\right] \period 
\label{eqn:aletap}
\eqaend
From the definition of $C_{-k}$, we get a compact result for the 
commutator  
\eqabegin
 \left[\alminus_m, C_{-k}\right] &=& {\gt\hbar \over i} C_{m-k}\period
\eqaend
With this result back in (\ref{eqn:aletap}), the commutator 
$\left[\alminus_m, \etap\right]$ is still quite involved due to 
 the presence of the inverse of $\calA$. What saves the day is 
 the fact that we only need to evaluate this operator between 
$\Omegabra$  and $\ket{\tilde{P}}$ both of which consist only of 
zero-modes. Thus effectively we can set all the non-zero 
 modes of $\psi$ ( denoted by $\tilde{\psi}$ ) 
to zero at this stage.  This greatly simplifies the rest of the 
calculation.  We easily check
\eqabegin
\left. C_k \right|_\psitz &=& \delta_{k,0} \comma \\
\left. \calA \right|_\psitz &=& \mu \e^{\psi^+_0}{1 \over 
\gt\pplus}\comma
\eqaend
and $\left[\alminus_m, \etap\right]$ takes an extremely simple form: 
\eqabegin
\left. \left[\alminus_m, \etap\right] \right|_\psitz
 &=& {\gt\hbar \over i} {\pplus \over \Pplus(m)}\e^{im\yplus}\period
\eqaend
Using this result, we get
\eqabegin
\left. \left[\alminus_m, \At_{-n}\right] \right|_\psitz &=& 
-{n\hbar \over \Pplus(m)} \Eplus{n}\alf_{m-n} \period\\
\eqaend
This leads to 
\eqabegin
\left. \left[G, \alminus_m\right] \right|_\psitz 
 &=& { 1\over \Pplus(m)} \sum_{n\ge 1} \tilde{\nu}_n \Eplus{n}
\alf_{m-n}\period
\eqaend
The double commutator $\left[G, \left[G,
\alminus_m\right]\right]$ can then be computed with the aid of the 
formula
\eqabegin
\left. \left[\alf_k, G\right] \right|_\psitz &=&
 \tilde{\nu}_k \Eplus{k} \qquad \mbox{(for $k\ge 1$ )} \\
 &=& 0 \qquad \qquad \mbox{( otherwise )} 
\eqaend
Putting together the results obtained so far, we get the 
re-ordering formula
\eqabegin
\left. \alminus_m \e^G \right|_\psitz 
&=& \e^G \left( \alminus_m -{1\over \Pplus(m)}\sum_{n\ge 1}
 \tilde{\nu}_n \Eplus{n} \alf_{m-n} \right. \nn\\
 & & -\left. \left. \half {1\over \Pplus(m)}\Eplus{m}\sum_{n=1}^{m-1}
 \tilde{\nu}_n\tilde{\nu}_{m-n} \right) \right|_\psitz  \period
\eqaend
To obtain $\Omegabra\e^{-\psi_0}\chia\e^G\ket{\tilde{P}}$, 
we multiply the above formula by $(\gt/im)\exp(-im\xplus)$ and 
sum over $m$ for $m\ge 1$.  Further, we can now set {\it all}
 the non-zero modes to zero.  The final result can be expressed in the 
 form
\eqabegin
&& \Omegabra\e^{-\psi_0}\chia\e^G\ket{\tilde{P}} \nn\\
&& \qquad =  i\gt\Omegabra \e^{-\psi_0} \left( \half \sum_{m\ge 1}
 {\Eplus{m} \over m\Pplus(m)}\sum_{n=0}^m \nu_n \nu_{m-n}
  \e^{-im(\xplus-x^+_0)}\right) \ket{\tilde{P}} \comma \\
\eqaend
where we have defined $\nu_0 \equiv p_f$ and extended 
the range of the sum over $n$ from $0$ to $m$ so as to incorporate 
 the term linear in $\nu_m$. Note that the phase factor included in 
$\tilde{\nu}_m$ has produced a shift in $\xplus$. \parsmallskip
%%%%%%%%%%%%%%%%%%%%%%%%%%%%%%%%%%%%%%
Calculation of $\calM_{1+}$ for $\calO=T^f$ proceeds 
 in a similar manner.  First the commutator is given by 
\eqabegin
 \left[\calLp, T^f\right] &=& \left[L^f_M, T^f\right] \nn\\
 &=& \hbar \e^{iM\xplus} \left\{ \left( \overi \delplus + 2M\right)
 T^f + \hbar 2M(M^2-1) \right\} \period
\eqaend
Thus we only need to compute $\Omegabra T^f \e^G \ket{\tilde{P}}$. 
The only non-trivial part is for $L^f_m \ (m\ge 1)$ in $T^f$.  Since 
$\ket{\tilde{P}}$ contains only zero-modes, we can replace 
$G$ by $G_f$ given by
\eqabegin
 G_f &\equiv& \left. G \right|_\psitz = 
\sum_{n\ge 1} {\tilde{\nu}_n \over \hbar n} \Eplus{n} \alf_{-n} \period
\eqaend
Therefore, 
\eqabegin
 \Omegabra L^f_m \e^G \ket{\tilde{P}} 
 &= &  \Omegabra \left[ L^f_m, \Gf\right]
 + \half \left[\left[ L^f_m, \Gf\right],\Gf\right]\ket{\tilde{P}}
\period
\eqaend
Commutators are easily evaluated. First using $\left[L^f_m, \alf_{-n}
\right] = n\hbar \alf_{m-n}$, we get
\eqabegin
 \left[L^f_m,\Gf\right] &=& \sum_{n\ge 1} \tilde{\nu}_n\Eplus{n}
\alf_{m-n} \period
\eqaend
Notice that this is very similar to 
$\left[G, \alminus_m\right]$ with $\psitz$ .  The double commutator 
 becomes
\eqabegin
  \left[\left[L^f_m, \Gf\right], \Gf\right]&=&
 \Eplus{m} \sum_{n=1}^{m-1} \tilde{\nu}_n \tilde{\nu}_{m-n} \period
\eqaend
Combining them we get 
\eqabegin
 \Omegabra L^f_m \e^G \ket{\tilde{P}} &=&
 \Omegabra \half \Eplus{m} \sum_{n=0}^m \tilde{\nu}_n \tilde{\nu}_{m-n}
 \ket{\tilde{P}} \period
\eqaend
With this formula we can easily compute the desired matrix 
 element as 
\eqabegin
&& \Omegabra\left[\calLp, T^f(\xplus)\right]\ket{\Psi_0} \nn\\
 &&\quad = -\hbar\e^{iM\xplus}\Omegabra\sum_{m\ge 0}(m-2M)
 \half \Eplus{m}\sum_{n=0}^m \nu_n\nu_{m-n}\e^{-im(\xplus-x^+_0)}
 \ket{\tilde{P}} \nn\\
 && \qquad + \hbar^2 \e^{iM\xplus}2M(M^2-1)\Omegabra\tilde{P}> \period
\eqaend
%%%%%%%%%%%%%%%%%%%%%%%%%%%%%%%%%%%%%
\medskip
The rest of the calculations are simpler than the ones sketched above 
 and we have  computed, without approximation, the mean 
values for  the operators $\del_{\xi^+}f$, $T^f(\xi^+)$, 
 $R^h_{+-}(\xi)=-\lambda^2\e^\psi$ and $g^{\alpha\beta}
 = -(\chi + AB)\e^{-\psi}\eta^{\alpha\beta}$. The results, which 
 are rather involved, are listed in  Appendix B. 
%%%%%%%%%%%%%%%%%%%%%%
\section{Black Hole Geometry in the Large $L$ Limit}
%%%%%%%%%%%%%%%%%%%%%%%%
\sectionnumbering
\sindent
In the previous section, we have perfomed a rigorous computation of
 the mean values for a number of operators of physical interest in a 
class of physical states. The results obtained are, however, still 
quite involved containing some infinite sums and are hard to interpret. 
In this section we shall evaluate these sums in the most interesting 
limit  where the (parameter) size of the universe, $L$, becomes very 
large. By choosing various parameters specifying the state 
appropriately, 
we  will be able to produce shock-wave-like energy-momentum 
distributions for the matter field and  see that in response black 
hole configurations will be formed.  \parsmallskip
%%%%%%%%%%%%%%%%%%%%%%%%%%
\subsection{ Preliminary Remarks }
%%%%%%%%%%%%%%%%%%%%%%%%%%
\sindent 
Before we begin our calculation, we must make several important remarks. 
\parsmallskipn
%%%%%%%%%%%%%%%%%%%%%%%%
$(i)$  The first remark is concerned with the precise meaning of the 
 large $L$ limit to be adopted in this article. Recall that to 
 make our analysis rigorous we have imposed spatially periodic 
 boundary conditions for the fields. 
This does not mean of course that our universe is necessarily 
homogeneous.  In fact we can arrange the parameters 
 so that the bulk of the matter energy-momentum density will be 
concentrated along a line $\xi^+ \sim \xi^+_0$ where 
 $\xi^+_0 = L x^+_0$.
 What we will look at is what happens in the 
 finite region around this line as $L \rightarrow \infty$. In other 
 words, our large $L$ limit is such that $x^\pm = \xi^\pm /L$ tend 
 to vanish.  ( This implies that we restrict ourselves to a finite 
 interval in the \lq\lq time variable" $\xi^0$ as well. As the 
 speed of light is finite, \ie unity in our convention, this is 
 causally reasonable. ) In this limit many of the terms in our 
 expressions of the mean values are easily seen to vanish.  
However, we must be very  careful in taking this limit for the terms 
involving the infinite sums. \parsmallskipn
%%%%%%%%%%%%%%%%%%%%%%%%%
$(ii)$  The second remark has to do with the phase factor 
 $e^{-ic p^+/\hbar \gamma}$ introduced  in the defintions of 
 $\ket{\Psizero}$ and $\Omegaket$.  ( See Eq.(\ref{eqn:psmearvac})
 and Eq.(\ref{eqn:Pkl}).)  This factor plays an important role 
 when the zero-mode $\qminus$ occurs in the mean value, as in 
 $< \nchiepsi >$: It produces a constant shift through 
$\e^{ic\pplus/\gamma\hbar} \qminus \e^{-ic\pplus/\gamma\hbar}
= \qminus +c/\gamma$ and 
 by varying $c$ we can  adjust the coordinate independent 
 contributions in $< \nchiepsi >$ freely. Thus in what follows the 
 coordinate-independent part of $<g^{\alpha\beta}>$ will be denoted 
 simply as an ajustable constant.
 \parsmallskipn
%%%%%%%%%%%%%%%%%%%%%%%%%
$(iii)$  Next we make a comment on the positivity of  $< T^f >$ 
 and the overall energy balance. The composite operator $T^f$ is 
 defined by a subtraction of an infinite positive constant and 
hence is not necessarily a positive definite operator.  One notices 
 however that the coordinate-independent part of $< T^f >$ proportional 
 to the zero-mode $p_f^2$ is positive and by a suitable choice of 
$p_f$ one can  always make $< T^f >$ positive throughout.  On the other 
hand, the coodinate-dependence of the metric is not affected by such 
 a choice.   
 This has a natural explanation coming from the energy-momentum 
constraint  $< T^{tot} > =0$ : It is easily checked that, 
for our choice of physical states, any change of $p_f$ 
 in $< T^f >$ is precisely compensated by the corresponding change in 
 the $< (1/\gt^2)\delplus\chi\delplus\psi >$ and consequently the 
 $<(1/\gt^2)\delplus^2\chi>$ part is left unchanged. This is 
responsible for the observed $p_f$-independence of 
 the coordinate-dependent part of $<\nchiepsi>$
(see Appendix B), which is the 
 only part of the metric that can possibly depend on $p_f$. In 
 physical terms, this phenomenon means that the positive uniform matter 
energy density is \lq\lq neutralized" by the negative uniform energy 
 density of the dilaton-Liouville sector. As the metric couples to 
 both of them, there is no net effect. \parsmallskipn
%%%%%%%%%%%%%%%%%%%%%%%%%%%%%%%%%%%%
$(iv)$ Finally, we make an important remark on the notion of 
\lq\lq quantum corrections".  A glance at the results listed in 
  Appendix B shows that all the coordinate dependence comes either
 with $\hbar^2/\kappa$ or with $\hbar^4/\kappa^2$ ( except for 
 a few terms which occur with an extra $\hbar$ ). Since $\kappa$ has 
 the dimension of $\hbar^2$, we shall henceforth set $\kappa = 
 \hbar^2$. Then, according to the usual terminology, bulk of the 
 contributions will become \lq\lq classical" and we have only a 
 few minor \lq\lq quantum corrections".  Is this a correct statement ? 
 The answer is interestingly ambiguous since in an exact quantum 
 treatment like the one we are pursuing the notion of \lq\lq quantum 
 corrections" becomes rather meaningless. We have chosen the state 
 $\ket{\Psizero}$ to be a coherent state so that we can expect to 
 produce \lq\lq classical" configurations. This is indeed achieved by 
 the  above assignment of $\kappa$. However, from the point of view 
 of the exact quantum theory, a coherent state is a highly 
 non-perturbative quantum state and precisely through its {\it quantum 
 coherence} \lq\lq classical objects" are formed.  This is 
 conceptually quite different from the semi-classical treatment where 
 strictly classical objects are provided from the beginning. 
 \parsmallskip
%%%%%%%%%%%%%%%%%%%%%%%%%%%%%%%%%%%
\subsection{ Large $L$ Limit }
%%%%%%%%%%%%%%%%%%%%%%%%%%%%%%%%%
\sindent 
With these remarks understood, we shall now take the large $L$ limit
 of the expressions listed in Appendix B. 
The infinite sums appearing for the coordinate-dependent parts are of 
 the generic form
\eqabegin
 S &=& a(L) \sum_{n\ge 1} b(n)c((n/L) \xi^+) \comma \nn\\
 &=& a(L)L \sum_{u=1/L, 2/L, \ldots} b(Lu)c(u\xi^+){1\over L} \comma
\eqaend
where we have introduced a variable $u\equiv n/L$ proportional to 
 the energy-momentum density. 
As $L$ becomes large,   this can be replaced by the integral 
\eqabegin
 S &\stackrel{L \rightarrow \infty}{\sim}& 
 a(L)L \int_{1/L}^\infty du b(Lu)c(u\xi^+) \comma
\eqaend
{\it provided} that the integral so obtained converges at both ends. 
\parsmallskip
%%%%%%%%%%%%%%%%%%%%%%%%
To examine this we now need to specify our parameters.  With the 
 purpose of producing shock-wave-black-hole configurations in mind, 
we have chosen them to be as follows:
\eqabegin
 \nu_n &=& \nu(Lu) = \nu u^d \e^{-au^2} \comma\\
 \omega_n &=& \omega(Lu) = {-\omega \over Lu} \qquad (n \ne 0) \comma\\
 \omega &=& \mbox{ a positive constant} \comma \\
 \omega_0 &=& \mbox{a constant to be adjusted } \comma\\
 \weight &=& \pplus \e^{-\alpha(\pplus-\pplus_0)^2/2} \comma
\eqaend
where $\nu$, $\omega$ and $\pplus_0$ are constants and we study 
 the cases for  $d= -1/2,\, 0,\, 1/2,\, 1$.  A factor of $\pplus$ 
in $\weight$
 is to suppress the contribution from $\pplus=0$ where various 
expressions become singular. \parsmallskip
%%%%%%%%%%%%%%%%%%%%%
Because of the Gaussian factor in 
 $\nu(Lu)$, the integrals are all convergent at the upper 
 end.  As for the lower end, one has to perform a somewhat tedious  
 power counting analysis.  The result of this analysis shows that 
 we can write the large $L$ limit of the mean values 
%( for $d\ge -1$ ) 
as 
\eqabegin
< \,\del_{\xi^+}f \, >&\stackrel{L \rightarrow \infty}{\sim}&
   \left(\frac{\gt p_f}{L}\, \calN 
  +2\gt \omega \nu I_f(\xiplus -\xiplus_0) \right) 
 \, \psmearnorm   \comma \\ 
%%%
<\,T^f (\xi^+) \,>&\stackrel{L \rightarrow \infty}{\sim}&
     \Biggl\{ \frac{ p^2_f}{2 L^2} \left( \calN + 6M^2(\omega^2 \zeta(2)
     + \omega^2_0) \right) 
       +\omega \nu^2 I_T(\xiplus -\xiplus_0 )\Biggr\}
  \, \psmearnorm    \comma \label{eqn:EXTf}\\
%%%%%%%%%%%%%
<\, R^h_{+-}(\xi)\,> &\stackrel{L \rightarrow \infty}{\sim}&
  -4\,\lambda^2  \,\Biggl[ \, \left( M\,\omega_0 +M^2(\omega^2\zeta(2)
 +\omega_0^2 )\right) \psmearnorm 
    + 2\gt^2 <\tilde{P} \mid p^2_+ \psmearvac \;\Biggr] 
 \nn \\
&=& \mbox{ constant } \comma \\
%%%%%%%%%%%
<\, g^{-1} \, > &\equiv & - <\, : \left( \chi+AB\right)\e^{-\psi} :\,> 
 \nn\\ 
 &\stackrel{L \rightarrow \infty}{\sim}&
\gt^2\omega\nu^2 I_\chi(\xiplus-\xiplus_0) \psmearnorm \nn\\
&&  -c_1\xiplus \delta_{d,-1/2}<\tilde{P}| \pplus \psmearvac - c_2 
\psmearnorm  \nn\\ 
& & -6\left({\lambda\over\gt}\right)^2 M^4
 \left(\omega^2 \zeta(2)+\omega_0^2\right)\xiplus\ximinus 
<\tilde{P} \mid \frac{1}{p^2_+}  \psmearvac \comma \label{eqn:ginv}
\eqaend
where  
 $I_f(\xi)$, $I_T(\xi)$ and $I_\chi(\xi)$ are integrals of the form 
\eqabegin
 I_f(\xi) &=& \int_{1/L} ^\infty du u^d \e^{-au^2}\cos u\xi \comma\\
 I_T(\xi) &=& \int_{1/L}^\infty du \cos u\xi 
\int_0^u dv \left[v(u-v)\right]^d \e^{-a(v^2 + (u-v)^2)} \comma 
 \label{eqn:ITdef} \\
 I_\chi(\xi) &=& \int_{1/L}^\infty du {\cos u\xi \over u^2}
\int_0^u dv \left[v(u-v)\right]^d \e^{-a(v^2 + (u-v)^2)} \comma
 \label{eqn:IKdef}  
\eqaend
and 
\eqabegin
 \calN &=& e^{\sum_{n \geq 1} \mid \nu_n \mid ^2 / \hbar n} \comma \\
 \zeta(2) &=& \mbox{ Riemann's $\zeta$ function } \comma \nn\\
< \tilde{P} \psmearvac &=&
   \left\{\; (p^+_0)^2 + \frac{1}{2 \alpha} \; \right\} \:
   \sqrt{\frac{\pi}{\alpha}}\;\gamma^4 \delta^{1+N_f} (0) \comma 
   \qquad (N_f = 24) \comma \\ 
< \tilde{P}\mid \pplus \psmearvac &=& \pplus_0 \left\{\,
 (\pplus_0)^2 + {3\over 2\alpha}\right\} \sqrt{\frac{\pi}{\alpha}}
\;\gamma^4 \delta^{1+N_f} (0)    \\
< \tilde{P} \mid  p^2_+ \psmearvac &=&
   \left\{\; (p^+_0)^4 + \frac{3}{\alpha} (p^+_0)^2 + \frac{3}{4\alpha ^2} 
 \; \right\}\: \sqrt{\frac{\pi}{\alpha}}\;\gamma^4 \delta^{1+N_f} (0),
   \comma  \\
< \tilde{P} \mid  \frac{1}{p^2_+} \psmearvac &=&
   \sqrt{\frac{\pi}{\alpha}}\;\gamma^4 \delta^{1+N_f} (0) \period
\eqaend
We must supply some explanations:
First, as it is an overall common factor, we did not bother to regularize
   $\delta^{1+N_f} (0)$ by additional smearing. Hence in the following, we
 shall consider quantities with $\psmearnorm$ removed. Secondly, 
 in bringing
 $<\, g^{-1}\, >$ to the above form, we have adjusted the constant 
$\omega_0$ in the following way. The large $L$ limit of 
$<\,AB\e^{-\psi}\,>$ originally contained a term of the form 
${\xiplus}^2+ {\ximinus}^2$ with a coefficient 
proportional to $\omega_0 -3M
 (\omega^2\zeta(2)+\omega_0^2)$. We have chosen $\omega_0$ to make this 
 coefficient vanish. This is a part of our gauge choice. 
Note that the term of the form $ -\lambda^2_g\xi^+\xi^- $ arises 
entirely from the pure gauge part of $<\,AB\e^{-\psi}\,>$ , and it 
describes the so called linear dilaton vacuum when the matter field
vanishes. The fact that our simple choice of $ \mid \Lambda > $ 
produces precisely such a vacuum configuration is quite remarkable.
Third comment is concerned with the third line of
 $<\, g^{-1}\, >$.  The term linear in $\xi^+$ with some 
 constant $c_1$ is an extra contribution present only for $d=-1/2$
 and it is obtained by a careful examination of the process of replacing 
 the infinite sums by integrals.  The magnitude of this term however 
 can be made arbitrary by changing  $\pplus_0$ and/or the 
exponent $\alpha$ in 
 the smearing function $\weight$.  The coordinate independent term 
 written as $c_2 \psmearnorm$, where $c_2$ depends on $p_f^2$,
 arises from several sources. But 
 this term can also be adjusted to any value following the remark $(ii)$ 
  above. 
%%%%%%%%%%%%%%%%%%%%%%%%%%%%%%%%%
\subsection{Evaluation of the Integrals }
%%%%%%%%%%%%%%%%%%%%%%%%%%%%%%%%%%
\sindent 
We now briefly describe how one can evaluate the integrals $I_f$, $I_T$
 and $I_\chi$. \parsmallskip
%%%%%%%%%%%
First, for $d>-1$ the integral $I_f$ can be expressed in terms of the 
 Kummer's confluent hypergeometric function $\Fii(a;b;z)$ as follows:
\eqabegin
 I_f &=& \half \Gamma\left( {d+1 \over 2}\right) a^{-(d+1)/2 }
 \Fii \left( {d+1 \over 2};\half; -{\xi^2\over 4a}\right) \\
&\stackrel{\xi\rightarrow 0}{\sim}&
 \half\Gamma\left( {d+1 \over 2}\right) a^{-(d+1)/2 }
 \left( 1-{d+1 \over 4a}\xi^2 + \calO(\xi^4) \right) \\
&\stackrel{\xi\rightarrow \infty}{\sim}&
2^d \sqrt{\pi}{\Gamma\left( {d+1\over 2} \right)\over 
 \Gamma\left( -{d\over 2}\right) } |\xi|^{-(d+1)} 
 \qquad ( d\ne 0,2,\ldots ) \period
\eqaend
For $d=0$, we have a simple Gaussian:
\eqabegin
 I_f(d=0) &=& \half \sqrt{\pi}a^{-1/2} \e^{-\xi^2/(4a)} \period \\
\eqaend
%%%%%%%%%%%%%%%%%%%%%
\indent 
$I_T$ and $I_\chi$ are of the similar type and can be treated
together as follows.
 First consider the integration over $v$. By making a 
change of variable from $v$ to $x$ of the form $x = 2(v-(u/2))/u$, 
 it can be transformed into 
\eqabegin
&& \int_0^u dv \left[v(u-v)\right]^d \e^{-a(v^2 + (u-v)^2)} \nn\\
 && \qquad = 2^{-2d}\e^{-au^2/2} u^{2d+1} \int_0^1 dx 
\left(1-x^2\right)^d \e^{-(a/2)x^2u^2} \period
\eqaend
Put this back into the integral $I_T$ or $I_\chi$ and interchange the 
 order of $u$- and $x$- integration. The result is 
\eqabegin
 I(d,\delta) &\equiv& 2^{-2d}\int_0^1 dx (1-x^2)^d
 \int_{1/L}^\infty du u^{2(d-\delta)+1}\e^{\alpha(x)u^2} \cos \xi u
 \comma \\
 \alpha(x) &\equiv & {a\over 2}(1+x^2) \comma
\eqaend
where $I_T = I(d,\delta=0)$ and $I_\chi = I(d,\delta=1)$. 
 For $d-\delta >-1$, $1/L$ can put to zero and the 
$u$-integration can be performed just like for $I_f$.
 In this way we obtain 
\eqabegin
 I(d,\delta) &=& 2^{-(d+\delta)}a^{-(d-\delta +1)}
 \Gamma(d-\delta +1) \nn\\
 & & \times \int_0^1 dx (1-x^2)^d (1+x^2)^{-(d-\delta+1)}
 \Fii \left( d-\delta +1; \half; {\xi^2 \over 2a(1+x^2)}\right)\period
\eqaend
Although the remaining $x$-integral cannot be performed in a closed 
form,  this is a very convenient form for numerical evaluation since 
 we no longer have oscillatory integral.   \parsmallskip
%%%%%%%%%%%%%
For $I_\chi$ for $d=-1/2\ \mbox{and}\ 0$,  the formula above does not 
apply since the integral diverges at the lower end 
 as $1/L$ tends to zero.  One can easily show, however, that 
 the divergent piece is a constant and hence inessential. ( Recall 
 the remark $(ii)$ again. ) After removing these pieces, 
the $u$-integral  can again be expressed in terms of confluent 
 hypergeometric functions.  \parsmallskip
%%%%%%%%%%%%%%%%%%%%%%%
 In Appendix C, we list the results 
 of all the relevant integrals together with their asymptotic behavior 
 for large $\xi$. ( For small $\xi$ they all behave like 
 $\sim A_0-A_2 (\xi^2/a) + \calO(\xi^4)$, where $A_0$ and $A_2$
 are  positive constants.) Utilizing these expressions and 
 numerical analysis thereof, we shall now be able to 
 give the physical interpretation of the final outcome of our long
 calculations. 
%%%%%%%%%%%%%%%%%%%%%%%%%%%%%%%
\subsection{ Physical Interpretation}
%%%%%%%%%%%%%%%%%%%%%%%%%%%%%%%%%
\sindent
A good starting point of our discussion is the examination of the 
 simplest of our results, namely  that of the 
 Ricci tensor $<\, R^h_{+-}(\xi)\, >= <\, -\lambda^2 \e^{-\psi}\, >
 = \mbox{constant} $. Although it says that 
with respect to $h_{\alpha\beta}$ our configurations are rather 
 trivial, it reveals an important nature of our choice of states.
 In the original work of CGHS and in many others which followed, 
 the gauge in which $\psi=0$  was recognized to be particularly 
 convenient for discussing various (semi-)classical black hole 
 configurations.  In such a gauge, one obviously has $R^h_{+-}
 = -\lambda^2= \mbox{constant}$, essentially of the same feature as 
 our $<\, R^h_{+-}(\xi)\, >$.  In our quantum calculation, this 
 comes about because our $\ket{\Psi}$ was designed to contain no 
non-zero modes of $\chi$ field:  Only the zero-mode part
 of $\e^{\psi}$ is active and hence no coordinate dependence arises 
 in $<\, R^h_{+-}(\xi)\, >$.  Thus our choice essentially corresponds 
 to the familiar gauge described above and this allows us to compare
 our results with the classical ones with ease.  \parsmallskip
%%%%%%%%%%%%%%%
Let us now describe  the results for the matter 
 energy-momentum tensor $<\, T^f(\xi) \,>$ and the inverse metric 
$<\, g^{-1}\,>$, obtained with the aid of  numerical calculations. 
It is convenient to denote the expression (\ref{eqn:ginv}) for 
 $<\, g^{-1}\,>$ in the following form:
\eqabegin
<\, g^{-1}\,> &=& -\kf(\xiplus) -\lm^2 \xiplus\ximinus \comma\label{eqn:EXg}
\eqaend
where
\eqabegin
 \kf(\xiplus) &=& c_K\xiplus + d_K -\gt^2\omega\nu^2 I_\chi(\xiplus
 -\xiplus_0) \comma \\
 c_K,\, d_K,\, \lm &=&\, \mbox{constants} \period
\eqaend
The constants can be adjusted as was discussed before, but we should 
 remember that $c_k$ can only be non-vanishing for $d=-1/2$ case. 
The curvature scalar is then given by 
\eqabegin
 R^g 
 &=& -4\lm^2 {\kf(\xiplus) -\xiplus \delplus \kf(\xiplus)
 \over \kf(\xiplus) + \lm^2 \xiplus\ximinus } \period\label{eqn:Rg}
\eqaend
Although we have performed numerical analysis for the four values 
 of $d$ which controles the behavior of the relevant integrals, we shall 
 only discuss $d=-1/2$ and $1/2$ cases in some detail 
 since the qualitative features for  the 
 remaining cases are not drastically different from these cases.
\parsmallskip
%%%%%%%
Let us begin with the $d=-1/2$ case. With suitable choice of 
 parameters, it describes precisely the space-time 
in which an in-falling ( smeared ) shock wave of matter energy produces 
a black hole without naked singularity, the prototypical configuration
 discovered in \cite{CG}.  From the expression given in  
(\ref{eqn:ITmhalf}), we see that the integral $I_T(\xiplus-\xiplus_0)$
 gives very nearly 
 a Gaussian peaked around $\xiplus_0$ and as the parameter $a$ 
 approaches $0$ it becomes a $\delta$-function.  As it is obviously 
 positive by itself, we set $p_f=0$ in the expression of $<\, T^f\,>$. 
( This is also convenient since for this value of $d$, the quantity 
$\calN$ diverges as $L$ becomes large. ) 
In Fig.1a we plot the typical behavior of $<\, T^f\,>$ and in Fig.1b
  the integral $I_\chi$ up to a constant. It is important 
 to note that the asymptotic behavior for large $|\xiplus -\xiplus_0|$ 
 is {\it linear} as is seen in (\ref{eqn:IKmhalf}).  Thus we can make 
 $<\, g^{-1}\,>$ behave very much like the CGHS case by adjusting 
 the term $c_K\xiplus $ in $\kf$, present for this value of $d$, and 
 the constant $d_K$, to cancel this linear portion for the range
$\xiplus < \xiplus_0$.   $\kf(\xiplus)$ then becomes
\eqabegin
 \kf(\xiplus) &=& \gt^2\omega \nu^2 \left( {\pi^2\over 2}
 (\xiplus -\xiplus_0) -I_\chi(\xiplus -\xiplus_0)\right) \comma
\label{eqn:kf}
\eqaend
which behaves like $ (\pi^2/2)\gt^2\omega \nu^2
\theta(\xiplus -\xiplus_0) $ for $|\xiplus -\xiplus_0|>>\sqrt{a}$. 
In this way we obtain a smeared version of the CGHS black hole. 
In Fig.1c, we show the line of curvature singularity for this 
configuration.  We clearly see that a black hole without a naked 
singularity is formed and to the left of the line $ \xi^+ = \xi^+_0 $ 
the space-time quickly becomes the linear dilaton vacuum configuration 
for $ \sqrt{a} << \xi^+_0 $ .
\parsmallskip
%%%%%
The total matter energy sent in is expressed in this \lq\lq Kruskal"
 coordinate system \\
as \cite{RS2}
\eqabegin
 E_f &=& \lm \int_0^\infty d\xiplus \xiplus <\,T^f(\xiplus)\,> \nn\\
  &=& \lm \omega \nu^2 \int_0^\infty d\xiplus \xiplus
 I_T(\xiplus -\xiplus_0) \period 
\eqaend
We may now use the important relation $I_T = -\del^2_{\xiplus}I_\chi$, 
 evident from the definitions (\ref{eqn:ITdef}) and 
 (\ref{eqn:IKdef}), which expresses the overall energy balance. 
Together with (\ref{eqn:kf}) above, this leads to 
\eqabegin
 E_f &=& {\lm\over \gt^2}\left.\left( \xi^+\del_{\xi^+}\kf -\kf\right)
 \right|_0^\infty \period
\eqaend
For sharply peaked matter distribution, this gives the familiar 
 result proportional to $\xiplus_0$:
\eqabegin
 E_f &=& {\pi^2\over 2}\lm\omega \nu^2 \xiplus_0 \period
\eqaend
The same expression can also be obtained from the point of view of 
 the energy stored in the dilaton-Liouville system. As our configuration
 is effectively a classical solution in the $\psi=0$ gauge, with 
$\lm$ as the dilatonic cosmological constant, it should be meaningful
 to look at the classical expression of the energy density $t_{00}$ of 
 the dilaton-Liouville system. It is straightforward to show that 
for $\psi=0$ it takes the form
\eqabegin
 t_{00} &=& \del_{\xi^1}^2 \Phi -2\lm^2 \comma 
\eqaend
which upon substituting $\Phi = -\lm^2\xiplus\ximinus -\kf$ becomes 
\eqabegin
 t_{00} &=& -\del_{\xi^1}^2\kf(\xiplus) = -\delplus^2 \kf(\xiplus)
 = t_{++} \period
\eqaend
Therefore the energy of the dilaton-Liouville system is obtained as 
\eqabegin
 E_{dL} &=& \lm \int_0^\infty d\xiplus\xiplus t_{++}(\xiplus) \nn\\
 &=& \left.\lm \left(\kf -\xiplus \delplus \kf \right)\right|_0^\infty
 \period
\eqaend
This is opposite in sign to the matter energy pumped in as expected. 
\parsmallskip
One might think that all these results indicate that we have simply reproduced a classical 
 cofiguration. This is not quite so: Comparison of $I_f$ with $I_T$ 
 given in (\ref{eqn:IFmhalf}) and (\ref{eqn:ITmhalf}) shows that 
 $<\, T^f\,>$ is not exactly  equal to the square of 
$<\,\del_{\xiplus} f\,>$. It is due to the fact that $T^f$ is 
 a composite operator and quantum interference has made the difference. 
In fact if {\it all} the expectation values behaved as classical, that 
would mean  that only a single state, which is a coherent state with 
 respect to all the operators, must dominate the intermediate sum. 
Such a situation is extremely difficult to arrange.  In the present
 context, the energy density, not $\del_{\xiplus} f$, is the 
operator of prime importance which directly influences the form of the 
 metric and hence we have arranged to make it behave as classically 
 expected. 
\parsmallskip
%%%%%%%%%%%%%%
Next let us consider the case for $ d=1/2 $. In this case, 
 $ I_T $ is given by an integral over a  Gaussian times 
a polynomial as shown in (\ref{eqn:IT12}). After the $x$-integral, it behaves 
like a Gaussian in the vicinity of  $ \xi^+ \sim \xi^+_0 $ , but 
develops a negative minimum and tends to vanish exponentially for large 
$ \mid \xi^+ - \xi^+_0 \mid $ .  Thus if we require the positivity 
of $ < T^f(\xi^+) > $ , we have to take $ p_f $ in (\ref{eqn:EXTf}) to be 
non-zero and of order $ L $.  As for  $ I_{\chi} $, it is almost 
a Gaussian as seen in  (\ref{eqn:IC12}) and we can easily find the behavior of 
$ < g^{-1} > $ by using (\ref{eqn:EXg}). In Fig.2, we plot the line of 
curvature singularity for this configuration, where we have taken 
 $ d_K $ to be an appropriate positive number and have chosen 
the width of the smeared shock wave to be rather broad.  One sees that 
while a space-like singularity is formed near $ \xi^+ \sim \xi^+_0 $, 
 as $\xiplus$ becomes large a time-like naked singularity develops 
and approaches the $ \xi^+ $ axis. This geometry is very similar to 
 the one that appears in the model of \cite{RS2}. 
\par
With non-vanishing  $ p_f $, the energy carried in by the matter field 
 becomes 
\eqabegin
E_f &=& \left.\frac{\lambda_g}{\tilde{\gamma}^2} ( \xi^+ \del_{\xi^+} 
K_g - K_g ) \right|^{\infty}_{0} + \,\lambda_g \frac{p^2_f}{2L^2}
 \left( \calN + 6M^2(\omega^2 \zeta(2) + \omega^2_0) \right) 
        \int^{\infty}_{0}d\xi^+\, \xi^+ \nn \\
    &=& {\rm negative \; constant } + \lambda_g \frac{p^2_f}{2L^2}
    \left( \calN + 6M^2(\omega^2 \zeta(2) + \omega^2_0) \right) 
       \int^{\infty}_{0}d\xi^+\, \xi^+ \period
\eqaend
Evidently the second term is divergent.  This was to be expected since 
 non-vanishing $ p_f^2 $ in $ < T^f > $ represents a 
 constant energy density permeating the whole universe 
 \cite{BG,RS2}. \parsmallskip
%%%%%%%%%%%%%%
Although the behavior of the line of curvature singularity resembles
 that of \cite{RS2}, there are several differences.  First, 
we do not impose additional boundary conditions on the naked 
singularity: The physical content  of our universe is 
already completely determined by the specification of the physical 
state.  Second, from (\ref{eqn:Rg}) we observe that the 
scalar curvature $ R^g $ goes to zero for large $\xiplus$, not to 
$ -\infty $ as in \cite{RS2}.  This means that the  black hole fades away 
in this region. 
%%%%%%%%%%%%%%%%%%%%%%%%
Finally, we comment on
the right-going part of the energy-momentum tensor of the matter fields,
$ < T^f_{--}(\xi^-) > $, the existence of which is often regarded as a 
signal of Hawking radiation . As our physical states contain only the 
left-movers in the non-zero mode sector and, due to conformal 
 invariance, they do not couple to the right-going counterparts, 
 only the zero-mode $ p_f $ contributes to $ < T^f_{--}(\xi^-) > $. 
After a simple calculation we get 
\eqabegin
< T^f_{--}(\xi^-) > &\stackrel{L \rightarrow \infty}{\sim}& 
         \frac{p^2_f}{2L^2} 
        \left( \calN + 6M^2(\omega^2 \zeta(2) + \omega^2_0) \right) 
         \psmearnorm \period
\eqaend
When $ p_f $ is of order $ L $, as in this case, 
$ < T^f_{--} > $ is finite, positive and coordinate-independent. 
There exsists no negative energy flow. This may be considered 
 as a rather general phenomenon: In the case of an unstable black hole 
 geometry, $ p_f $ should be non-zero from the requirement of 
the positivity of $ < T^f_{++} > $ , and this in turn  leads to 
finite and positive $ < T^f_{--} > $.  Further discussion  on the 
relation between Hawking radiation and $ < T^f_{--} > $, however, 
 is unfortunately beyond the scope of this paper. 
%%%%%%%%%%%%%%%%%%%%%%%%%
\section{Discussions}
%%%%%%%%%%%%%%%%%%%%%%%%
\sindent
By using an exactly solvable model of CGHS type, we have shown 
explicitly how one can extract space-time geometry from an 
exact yet abstract physical state in quantum theory of gravity. 
Although the model employed is but a toy model in $1+1$ dimensions, 
we believe that it is quite significant to be able to discuss a variety of 
 important issues in quantum gravity in a concrete and unambiguous manner. 
Furthermore, the point of view and the procedures developed 
in this work should find wide applications in other models of 
quantum gravity and possibly in quantum cosmology. \parsmallskip
%%%%%%%%%
In spite of the progress made, a number of 
important problems still remain to be understood. 
The first and the foremost is the question of how to define and 
compute the S-matrix: Without its understanding, we cannot even 
formulate the most interesting problem concerning the Hawking 
radiation, quantum coherence, and the fate of a black hole. 
\parsmallskip
%%%%%%
There appear to be two substantial obstacles we must overcome. 
In semi-classical treatment, a reference background geometry 
as well as a coordinate system are already available, and  
one has a space-time picture  before
 one starts discussing the S-matrix for particle excitations 
around such a macroscopic background.  On the other hand, in 
exact analysis geometry can emerge only {\it after} we compute 
 some expectation values of appropriate operators, as we have seen 
in this work. In other words, it is extremely difficult to separate 
out the bulk geometry and the particle excitations  for which 
to define the scattering matrix. \parsmallskip
%%%%%%
 The second difficulty has to do with the very nature of the usual 
 definition of the S-matrix.  S-matrix elements describe the overlap 
between the \lq\lq states " prepared in the \lq\lq far past " 
and the ones defined in the \lq\lq far future ", where  
 the interactions are supposed to be negligible. 
Thus the notion of S-matrix inherently hinges upon our ability to 
separate out appropriate sub-regions or sub-systems. 
 This is again very hard to 
do in advance {\it before} we obtain a space-time picture, 
in particular before the notion of \lq\lq time " becomes available. 
To avoid any confusion, 
we emphasize that it is not the question of \lq\lq boundary 
conditions " as sometimes argued. 
Such a terminology implicitly assumes that one already 
has a space-time picture, which we do not. Besides,  boundary 
conditions are already used in obtaining the physical states and 
are not to be imposed again after the extraction of a space-time 
picture. In any case, the pressing task is to give a concrete 
 procedural definition of the S-matrix in quantum gravity 
 to which everyone can agree. \parsmallskip
%%%%%%%%%%%%%%%%%%%%%%
Finally, as for the  Hawking radiation and the problem of loss of quantum 
coherence, we do not have much to say since one cannot 
make a proper discussion of this issue without a satisfactory definition of 
the S-matrix. Nevertheless it may be worth remarking that an 
essentially similar situation is expected to occur in a many-body system,
 even without gravity, 
where the system can be approximately divided into a macroscopic classical
 sub-system  and a microscopic quantum sub-system. If one is able 
 to treat the whole system exactly, there should not be any quantum 
incoherence. On the other hand, if one makes an approximation 
 as stated above, the macroscopic part would  act as a germ 
 of incoherence for the quantum sub-system.  It would be quite 
interesting if one can set up a simple idealized  model in which to 
study the emergence of quantum incoherence in an explicit manner. 
%%%%%%%%%%%%%%%%%%%%%%%%%%%%%
\parbigskipn\parbigskipn
 {\Large\bf Acknowledgment } \parbigskip
We would like to thank S. Hirano for his  contribution concerning
 the DDF construction. Y.K. thanks M. Kato and A. Shimizu  
 for a discussion, respectively, on the issue of the inner product
 and on the problem of  quantum coherence, while 
Y.S. acknowledges useful conversations with 
 K. Hori, I. Ichinose, A. Tsuchiya and in particular H. Ishikawa.  
 The research of Y.K. is supported in part 
by the Grant-in-Aid for Scientific Research (No.04640283) 
 and Grant-in-Aid for scientific Research for Priority Areas  
(No. 05230011) from the Ministry of Education,Science and Culture.
%%%%%%%%%%%%%%%%%%%%%%%%%%%%%%%%%%%
%%%%%%%%%%%%%%%%%%%
\appendixnumbering{A}
\parbigskipn\parbigskipn
{\Large\bf Appendix A} \parbigskip
%%%%
In this appendix, we shall derive an explicit form of the new 
field $\eta^+$ introduced in Eq.(\ref{eqn:etap}).  Furthermore, 
we define its 
conjugate field $\zeta^+$ such that at the classical level 
 the transformation from $(\chi^+, \psi^+)$ into $(\eta^+, \zeta^+)$ 
 is a canonical transformation.  Despite the non-local nature of 
this transformation, the energy-momentum tensor in the dilaton-Liouville 
sector will be seen to take a simple local form in terms of the new 
 pair of fields. 
%%%%
\parag{Explicit Expression of $\etap$ }
%%%%
\indent
We begin by deriving a useful alternative representation of 
  $\calA(\xplus)$.  First it is clear from Eq.(\ref{eqn:calA}) that 
 $\qplus$-dependence of $\calA(\xplus)$ is simply an overall factor 
$\exp(\gt\qplus)$.  Next due to the boundary condition 
$\calA(\xplus + 2\pi) = \alpha \calA(\xplus)$ with 
$\alpha = \exp(\gamma\sqrt{\pi}\pplus)$, $\calA(\xplus)$ must contain a 
factor $\exp(\gt \pplus \xplus )$ and the rest must 
be periodic. This means that $\calA(\xplus)$ can be written as 
\eqabegin
 \calA(\xplus) &=& \mu \calA_0(\xplus) \tilde{\calA}(\xplus) \comma \\
 \calA_0(\xplus) &=& \e^{\gt(\qplus + \pplus \xplus)}
 = \e^{\psi^+_0} \comma \\
 \tilde{\calA}(\xplus + 2\pi) &=& \tilde{\calA}(\xplus) \period
\eqaend
By applying $\delplus$ on $\calA(\xplus)$ above and comparing it with
 the original definition of $\calA(\xplus)$ 
\ie $\delplus \calA = \mu \e^{\psi^+}$, we easily deduce 
\eqabegin
 \e^{\tilde{\psi}^+} &=& \left(\delplus + {\pplus 
 \over \sqrt{2}Q } \right) \tilde{\calA} \\ 
 &\equiv & \hat{{\cal P}}_+ \tilde{\calA} \comma 
\eqaend
where $Q=\sqrt{2\pi}/\gamma$ is the background charge. In the space 
 of periodic functions the differential operator $\hat{{\cal P}}_+ $ 
 can be  inverted. Indeed by Fourier analysis, we can easily obtain the 
Green's function $g(x^+ -y^+)$ for it as 
\eqabegin
 \hat{{\cal P}}_+  g(x^+ -y^+) &=& 2\pi \delta(x^+ -y^+) \\
 g(x^+-y^+) &=& \sqrt{2} Q \sum_{m\in {\bf Z}}
 {1\over \Pplus(m)}\e^{im(x^+-y^+)} \comma
\eqaend
where $\delta(x^+-y^+)$ is the periodic $\delta$ function and 
$\Pplus(m)$ is precisely the quantity that appeared in the construction
 of the BRST cohomology by means of the operator $T^+$ ( see Eq.
(\ref {eqn:Kplus}) ).
 Thus $\tilde{\calA}$ can be solved to be 
\eqabegin
 \tilde{\calA} &=& \int_0^{2\pi} {d\yplus \over 2\pi} g(\xplus -\yplus)
 \e^{\tilde{\psi}^+(\yplus)} \\
&=& \sqrt{2} Q \sum_{m\in {\bf Z}} {1\over \Pplus(m)}\e^{imx^+}
 C_{-m} \comma \\
 C_{-m} & \equiv & \int_0^{2\pi} {d\yplus \over 2\pi}\e^{-im\yplus}
 \e^{\tilde{\psi}^+(\yplus)} \period
\eqaend
Thus we obtain 
\eqabegin
 \calA(\xplus)/\mu  &=&  \e^{\psi^+_0}\cdot 
 \sqrt{2} Q \sum_{m\in {\bf Z}} {1\over \Pplus(m)}\e^{imx^+}
 C_{-m} \period
\eqaend
Taking the logarithm of this expression and separating the 
result into the zero-mode and the non-zero-mode parts, we get
\eqabegin
 \eta^+ &=& \eta^+_0 + \tilde{\eta}^+ \comma \\
 \eta_0^+ &=& \psi^+_0 + \ln\left( {C_0 \over \gt\pplus}
 \right)\comma \\
 \tilde{\eta}^+ &=& \ln \left( 1+ \sum_{n\ne 0}{\pplus \over 
\Pplus (n)}{C_{-n} \over C_0} \e^{in\xplus} \right) \period
\eqaend
Note that  the zero-mode $\etazp$ contains, apart from the zero-mode
 part of $\psi^+$, the expression $\ln\left( 
 C_0/ \gt \pplus \right)$  which is a complicated 
combination of non-zero modes of $\psi^+$.  \parsmallskip
The inverse relation, \ie the expression of $\psip$ in terms of 
$\etap$, is immediately obtained from the defining relation 
$\delplus \e^{\etap} = \e^{\psip}$ as
\eqabegin
 \psip &=& \etap + \ln(\delplus\etap) \period \label{eqn:psip}
\eqaend
%%%%%%%%%%%%%%%%%%%
\parag{Classical Canonical Transformation }
%%%%%%%%%%%%%%%%%%%
\indent
We shall now try to find a field $\zetap$ the modes of which are 
conjugate to those of  $\etap$, and express the energy-momentum tensor 
in terms of new fields $\etap$ and $\zetap$. \parmedskip
To find $\zetap$, we shall perform a classical canonical transformation
 from the \lq\lq old" set of canonical pair $(q,p) \sim (\chip, \psip)$ 
to 
 the \lq\lq new" set $(Q,P) \sim (\etap, \zetap)$.  To be rigourous, 
let us look at the individual modes and identify the canonical pair:
\eqabegin
&& (q,p) \quad \sim \quad ( \qminus,\pplus),\ (\pminus, -\qplus), \
 ( \alminus_n /n, i\alplus_{-n}) \nn\\
&& (Q,P) \quad \sim \quad (\qe,\pz),\ (\pe, -\qz),\ 
 (-\beta^+_{-n}/n, i\beta^-_n ) \nn
 \eqaend
where we denote the non-zero modes of $\etap$ and $\zetap$ by $\beta^+_n$
 and $\beta^-_n$ respectively. 
Our convention for the Poisson bracket is
$$ \left\{q,p \right\}=1, \qquad \left\{\alpha^+_n, \alpha^-_{-n}
 \right\} = {1\over i} n \period $$
We take the generating function to be of type $F(q,Q)$.  Then, as is 
well known, 
\eqabegin
 p&=& {\del F \over \del q}, \qquad P = -{\del F \over \del Q} \period
\eqaend
To realize the first of these relations, we must have
\eqabegin
 F&=& \qminus \pplus -\pminus \qplus + \sum_{n\ne 0}{i\over n}
 \alminus_n \alplus_{-n} \period
\eqaend
The rest of the procedure is to express $\pplus, \qplus, \alplus_{-n}$ 
in terms of  $\qe, \pe, \beta^+_n$ and then vary $\qe, \pe, \beta^+_n$ 
to get the modes of $\zetap$. \parsmallskip
%%%%%%%%%%%%%%%%%%%%%%%%%%%%
Making use of the explicit form of $\etap$, we obtain, after 
 some amount of work, the following expressions:
\eqabegin
 \qz &=& \qminus -\pminus \xpint {1\over \delplus \etap} \comma\\
 \pz &=& \pminus \comma\\
 \beta^-_n &=& in\pminus \ypint {1\over \delplus \etap}\e^{in\yplus}
 \nn\\
 & & + \alminus_n + in {\sqfp\over \gamma} 
 \ypint {\delplus\chitp\over \delplus
 \etap} \e^{in\yplus} \period
\eqaend
We can now form the field $\zetap$ in the usual way:
\eqabegin
 \zetap(\xplus) &=& {\gamma \over \sqrt{4\pi}}
 \left( \qz + \pz \xplus  + i \sum_{n\ne 0}{\beta^-_n \over n} 
\e^{-in\xplus}\right) \period
\eqaend
Using the identity $\sum_{n \ne 0} \exp (in(\yplus -\xplus)) =
 2\pi \delta (\xplus -\yplus) -1$, we get
\eqabegin
 \zetap &=& \chip -{\delplus \chip \over \delplus \etap} 
 + \ypint {\delplus \chitp \over \delplus \etap} \period
 \label{eqn:zetap}
\eqaend
Since the last integral is independent of $\xplus$, the derivative 
$\delplus\zetap$ takes the form
\eqabegin
 \delplus \zetap &=& \delplus \chip -\delplus 
\left( {\delplus \chip \over \delplus \etap} \right) \nn\\
 &=& {1\over \delplus \etap} \left[ \delplus \chip \left( \delplus
 \etap + {\delplus^2 \etap\over \delplus \etap }\right) -\delplus^2 \chip
 \right] \period
\eqaend
%%%%%%%%%%%%%%%%%%%%%%%%%%%%%%%
If we write the last equation in the form 
\eqabegin
 \delplus \zetap \delplus \etap 
 &=& \delplus \chip \left( \delplus
 \etap + {\delplus^2\etap \over \delplus \etap} \right) -\delplus^2 
\chip \comma 
\eqaend
and use the relation
\eqabegin
\delplus\psip &=& \delplus \etap + {\delplus^2 \etap
\over \delplus \etap } \comma
\eqaend
which follow from (\ref{eqn:psip}), 
we recognize that the above expression is nothing but 
the energy-momentum tensor in the dilaton-Liouville sector. Thus 
we obtain
\eqabegin
 {\gt}^2 T^{dL}(\xplus) &=& \delplus\psip \delplus\chip 
 -\delplus^2\chip  \comma\\
  &=& \delplus\zetap \delplus \etap \period
\eqaend
It is remarkable that we have obtained a simple local expression
 of free-field type {\it without a background charge} 
despite the  complicated non-linear and non-local transformations 
perfomed. One recognizes that $\etap$ and $\zetap$ are essentially 
 the fields employed in the work of \cite{VV1,VV2}. However, 
as the form of $\zetap$ ( Eq.(\ref{eqn:zetap}) ) shows, 
the canonical relation between $(\chip, \psip)$ and $(\etap, \zetap)$ 
cannot be easily extended to the quantum domain. 
%%%%%%%%%%%%%%%%%%%%%%%%%%%%%%%%
%%%%%%%%%%%%%%%%%%%
\newpage
\appendixnumbering{B}
\parbigskipn\parbigskipn
{\Large\bf Appendix B} \parbigskip
%%%%%%%%%%%%%%%%%%%%%%%%%%%%
In this appendix, we list the exact results of the calculations of the 
 mean values for the operators $\del_{\xi^+}f$, $T^f(\xi^+)$, 
 $R^h_{+-}(\xi)=-\lambda^2\e^\psi$ and $g^{\alpha\beta}
 = -(\chi + AB)\e^{-\psi}\eta^{\alpha\beta}$ before we take the 
 large $L$ limit. 
%%%%%%%%%%%%%%%%%%%%%%%%%%%%%%%%
\eqabegin
 <\del_{\xi^+}f\, > &=& (F_1+F_2+F_3) \psmearnorm 
 \phantom{ \mbox{ \hspace{5cm} }}
\eqaend
\eqabegin
F_1 &=& \frac{\gt p_f}{L}\, 
    \exp\left(\sum_{n \geq 1} {\mid  \nu_n \mid ^2 \over \hbar n}\right)
     \comma  \\
F_2 &=& -2 \,\frac{\gt \hbar^2}{\kappa L} 
           \sum_{n\geq 0} \,\nu_n \,\omega_n\:(n-M)\,
      \cos\,(n(x^+ - x^+_0)-Mx^+)
              \comma  \\
F_3 &=& -12\,\frac{\gt \hbar^4}{\kappa^2 L} M p_f 
    \sum_k k \omega_k^2
\eqaend
%%%%%%%%%%%%%%%%%%%%%%%%%%%%%%%
\eqabegin
< T^f(\xi^+) \,> &=& (T_1+T_2+T_3) \psmearnorm \comma 
 \phantom{ \mbox{ \hspace{5cm} }}
\eqaend
\eqabegin
T_1 &=& \frac{ p^2_f}{2 L^2} 
       \exp\left(\sum_{n \geq 1} {\mid  \nu_n \mid ^2 \over
  \hbar n } \right)        \comma  \\
T_2 &=&  2\frac{\hbar^2}{\kappa L^2} \left\{
     \:\cos\,Mx^+ \{Mp_f^2 +
     \frac{\hbar}{12}\,c_f\,M(M^2 - 1) \} \;\omega_0 \right. \nn \\
    && \left. + \sum_{k \geq 1} \omega_k \,\{\,p_f \nu_k + \half 
   \sum_{n=1}^{k-1}\, \nu_n \nu_{k-n} \} \,(2M-k)\,
   \cos\,( k(x^+-x^+_0)-Mx^+ )
      \right\}  \comma \\ 
T_3 &=& {\hbar^4\over \kappa^2 L^2}\left\{ -6M p_f^2
    \,\sum_k k\omega^2_k \right.    \nn \\
 && \left. + 3 \left[M^2 p_f^2 +
    \frac{\hbar}{12}\,c_f\,M^2(M^2 - 1) \right]
 \,\sum_k \omega_k^2 \right\}  \period
\eqaend
%%%%%%%%%%%%%%%%%%%%%%
\eqabegin
 < R^h_{+-}(\xi) \, > &=& < -\lambda^2 \e^{\psi} \,> 
   = R_1+R_2+R_3 \comma 
 \phantom{ \mbox{ \hspace{5cm} }}
\eqaend
\eqabegin
R_1 &=& -2\lambda^2{\hbar^2\over \kappa}\omega_0
<\tilde{P} \mid \left(\,\gt p^+ \,
(\sin\,Mx^+ + \sin\,Mx^-) \right. \nn\\
  && \left.+  M\,(\cos\,Mx^+ +\cos\,Mx^-)\right)\,
   e^{ 2\gt p^+ t} \psmearvac \comma\\
R_2 &=& 8\lambda^2 M {\hbar^4\over \kappa^2}\sum_k k \omega_k^2 
 <\tilde{P} \mid e^{ 2\gt p^+ t} 
  \psmearvac  \nn \\
 & & -4\lambda^2{\hbar^4\over \kappa^2}
\,\sum_k \omega_k^2
   <\tilde{P} \mid (\gt p^+ )^2 
        \,e^{ 2\gt p^+ t} \psmearvac  \comma\\
R_3 &=& \!\!-4\,\lambda^2 \frac{\hbar^4}{\kappa^2} 
             (\cos\,Mx^+ \cos\,Mx^- +\sin\,Mx^+\sin\,Mx^-) \nn\\
 & & \times \sum_k \omega_k^2
       <\tilde{P} \mid (\gt p_+)^2 + M^2 \psmearvac
 \period
\eqaend
%%%%%%%%%%%%%%%%%%%%%%%%%%%%%%%%%%%%
\eqabegin
< \, AB\,e^{- \psi} \, > &=& \left((AB)_1 +(AB)_2 + (AB)_3+ (AB)_4 
 \right) <\tilde{P} \mid \frac{1}{p^2_+} \psmearvac 
 \phantom{ \mbox{ \hspace{2cm} }}
\eqaend
\eqabegin
(AB)_1 &=& {\lambda^2 L^2\over \gt^2} 
      \exp\left(\sum_{n \geq 1} {\mid  \nu_n \mid ^2 \over
  \hbar n } \right)         \comma \\
(AB)_2 &=& -2{\lambda^2 L^2\over \gt^2 }{\hbar^2 \over \kappa}M\omega_0
 \left( \cos\, Mx^+ + \cos\, Mx^- \right) 
    \comma \\
(AB)_3 &=& 12{\lambda^2 L^2\over \gt^2 }{\hbar^4\over \kappa^2}
   M \sum_k (M-k)\omega_k^2 \, 
    \comma \\
(AB)_4 &=& 6M^2{\lambda^2 L^2\over \gt^2 }{\hbar^4\over \kappa^2}
  ( \cos\,Mx^+\cos\,Mx^- + \sin\,Mx^+\sin\,Mx^-) 
    \period 
\eqaend
%%%%%%%%%%%%%%%%%%%%%%%%%%%%%%%%%%%%%%
\eqabegin
 <\, :\chi \,e^{- \psi}: \,> &=& C_{11}^+ +C_{11}^- +C_{12}^+
 + C_{12}^- +C_{21}+C_{22}+C_3 
  \comma 
 \phantom{ \mbox{ \hspace{5cm} }}
\eqaend
\eqabegin
C_{11}^+ &=& - \frac{2 \hbar^2}{ \kappa} \Bigl\{ \,
    - \gt^2\hbar M\,\cos\,Mx^+ 
  <\Psi_0 \mid \,e^{-\gt (2p^+t + q^+)} 
  \omegavac \nn \\
&& \quad + ( M\,\cos\,Mx^+ - \sin\,Mx^+\, \del_+) \nn\\
&& \qquad \cdot <\Psi_0 \mid \gt (2p^- t + (c/\gamma) )
   \, e^{-\gt (2p^+t + q^+)} \omegavac \nn \\
 && \quad -( M\,\sin\,Mx^+ + \cos\,Mx^+\, \del_+) \nn\\
&& \qquad \cdot <\Psi_0 \mid \gt i q^- \, e^{-\gt (2p^+t + q^+)}
 \omegavac  \Bigr\} \comma \\ 
C_{11}^- &=& C_{11}^+(x^+ \rightarrow x^-, \del_+ \rightarrow \del_- ) 
 \comma \\
C_{12}^+ &=& - 2\gt {\hbar^2 \over \kappa}\,
    <\tilde{P} \mid \,e^{-2\gt p^+t} \Biggl[\:
      p_f \,\sum_{n \geq 1} (n+M)  
    \frac{\nu_n \omega_n}{n(p_+^2+ 2Q^2 n^2)} \nn\\
&& \cdot \Bigl\{ -\,\cos\,Mx^+ \;\left(p^+ \sin\,n(x^+-x^+_0)+
  \sqrt{2}Qn\, \cos\,n(x^+-x^+_0)\right) \nn\\
&& +\,\sin\,Mx^+\left(p^+ \cos\,n(x^+ -x^+_0)
  - \sqrt{2}Qn\,\sin\,n(x^+-x^+_0)\right) \Bigr\}  \nn \\
& & + \half \,\sum_{m,n \geq 1} (m+n+M)  
    \frac{\nu_m\nu_n \omega_{m+n}}{(m+n)(p_+^2+ 2Q^2 (m+n)^2)}  \\
&&  \cdot  \Bigl\{ -\,\cos\,Mx^+ \; \left(p^+ \sin(m+n)(x^+ -x^+_0)
 +  \sqrt{2}Q(m+n) \cos(m+n)(x^+-x^+_0)\right) \nn \\
&&  +\sin\,Mx^+ \Bigl( p^+ \cos(m+n)(x^+-x^+_0) \nn\\
&& - \sqrt{2}Q(m+n)\sin(m+n)(x^+-x^+_0) \Bigr) \Bigr\} \; \Biggr] 
      \psmearvac  \comma  \\
C_{12}^- &=& - 2\gt {\hbar^2 \over \kappa}\,
    <\tilde{P} \mid \,e^{-2\gt p^+t} \Biggl[\:
      p_f  M\,\sum_{n \geq 1}  
    \frac{\nu_n \omega_n}{n(p_+^2+ 2Q^2 n^2)} \nn\\
&& \cdot \Bigl\{ -\,\cos\,Mx^+ \;\left(p^+ \sin\,n(x^+-x^+_0)+
  \sqrt{2}Qn\, \cos\,n(x^+-x^+_0)\right) \nn\\
&& +\,\sin\,Mx^+\left(p^+ \cos\,n(x^+ -x^+_0)
  - \sqrt{2}Qn\,\sin\,n(x^+-x^+_0)\right) \Bigr\}  \nn \\
& & + \half  M \,\sum_{m,n \geq 1} 
    \frac{\nu_m\nu_n \omega_{m+n}}{(m+n)(p_+^2+ 2Q^2 (m+n)^2)}  \\
&&  \cdot  \Bigl\{ -\,\cos\,Mx^+ \; \left(p^+ \sin(m+n)(x^+ -x^+_0)
 +  \sqrt{2}Q(m+n) \cos(m+n)(x^+-x^+_0)\right) \nn \\
&&  +\sin\,Mx^+\Bigl(p^+ \cos(m+n)(x^+-x^+_0)\nn\\
&& - \sqrt{2}Q(m+n)\sin(m+n)(x^+-x^+_0) \Bigr) \Bigr\}\;\Biggr] 
      \psmearvac  \comma  \\
C_{21} &=& {\rm Re} \left\{ 4\,M \frac{\hbar^4}{\kappa^2} <\Omega \mid 
     \gt (2p^- t + q^-)\, e^{ -\gt 
    (2p^+ t + q^+)}\,(L_0/\hbar)  \omegavac \right\}  \\ 
C_{22} &=& 
    {\rm Re}\Biggl\{ 2\frac{\hbar^4}{\kappa^2} [\; - <\Omega \mid 
  \left\{ \:  ( 2\gt p^+ +iM )
      \gt p^- + 3\hbar \gt^2 M^2 \right\}
      \,e^{ -\gt (2p^+ t + q^+)} 
     \omegavac  \nn \\
&&+   <\Omega \mid \gt (p^- t + q^-)
       ( \;(\gt p^+)^2 +iM\gt p^+  +2 \,M^2)
      \, e^{ -\gt (2p^+ t + q^+)} \omegavac
      \; ] \Biggr\}  \\
C_3 &=& {\rm Re}\Biggl\{ 2\frac{\hbar^4}{\kappa^2}
     \,e^{iMx^+}\,e^{-iMx^-} \nn\\
 && \cdot <\Omega \mid \,\Bigl[\;
     -2\gt^2(\hbar M^2 + p^-p^+) 
+ \gt (2p^- t + q^- )\left(M^2 + 
     (\gt p^+)^2 \right)
    \;\Bigr] \nn\\
& & \cdot e^{-\gt(2p^+t + q^+ )} \omegavac \Biggr\}\period
\eqaend
%%%%%%%%%%%%%%%%%%%%%%%%%%%%%%%%%%%%%%%%%
In the expressions above, some of the quantities in the final stage
 of the calculation are left unevaluated.  This is because of the 
 following two reasons:  These quantities depend on 
 how we choose the smearing function $\weight$ and should better be 
evaluated after we specify $\weight$.  Moreover, many of them actually 
 vanish as $L\rightarrow \infty$, the limit we are most interested in. 
%%%%%%%%%%%%%%%%%%%
\newpage
\appendixnumbering{C}
\parbigskipn\parbigskipn
{\Large\bf Appendix C} \parbigskip
%%%%%%%%%%%%%%%%%%%%%%%%%%%%
In this appendix, we provide a list of relevant integrals that 
 occur in the calculation of the mean values and their large $\xi$ 
 asymptotic behavior.  For certain special cases, the confluent 
 hypergeometric function reduces to ( a polynomial times ) a Gaussian 
 and we shall use such a simplified form whenever it occurs.  For 
 such essentially Gaussian cases, large $\xi$ form will not be 
 listed. 
%
%%%%%%%%%%%%%%%%%%%%%%%%%%%%%%%%%%%%%%%%%%%%%%%%%%%
\parag{$d=-1/2$ case }
%%%%%%%%%%
\eqabegin
 I_T &=& \sqrt{{2\pi\over a}}\int_0^1 dx (1-x^2)^{-1/2}(1+x^2)^{-1/2}
 \e^{-\xi^2/(2a(1+x^2))} \label{eqn:ITmhalf} \\
 I_\chi &=& L\pi  -\sqrt{2\pi a}\int_0^1 dx
\cx{-1/2}{1/2} \e^{-\xi^2/(2a(1+x^2))} \nn\\
 && \quad -\sqrt{{2\pi \over a}}\xi^2 \int_0^1 dx
 \cx{-1/2}{-1/2} \Fii\left( \half; {3\over 2};
 -{\xi^2 \over 2a(1+x^2)} \right)  \\
\xilarge L\pi -{\pi^2\over 2}|\xi| \label{eqn:IKmhalf} \\
 I_f &=& \half \Gamma\left({1\over 4}\right) a^{-1/4}\, 
\Fii\left( {1\over 4};\half;-{\xi^2 \over 4a}\right) 
\label{eqn:IFmhalf}\\
\xilarge \sqrt{{\pi \over 2}}\, |\xi|^{-1/2} 
\eqaend
%%%%%%%%%%%%%%%
\parag{$d=0$ case }
%%%%%%%%%%%%%%%%%%
\eqabegin
I_T &=& {1\over a} \int_0^1 dx {1\over 1+x^2} \Fiialpha{1}{\half} \\
\xilarge -\xi^{-2} \\
I_\chi &=& \int_0^1 dx \int_{1/L}^\infty {du \over u}
 \e^{-\alpha(x)u^2} \nn\\
 & & \quad - {1\over a}\int_0^\xi d\zeta \zeta \int_0^1 dx
 {1\over 1+x^2} \Fii\left( 1;{3\over 2};-{\zeta^2\over 2a(1+x^2)}\right)
\\
\xilarge const. -\ln \xi \\
I_f &=& \half \sqrt{\pi} a^{-1/2} \e^{-\xi^2/(4a)} 
\eqaend
%%%%%%%%%%%%%%
\parag{$d=1/2$ case }
%%%%%%%%%%%%%%
\eqabegin
I_T  &=& 2^{-3/2}\sqrt{\pi}a^{-3/2} \int_0^1 dx \cx{1/2}{-3/2}
 \left( 1-{\xi^2 \over a(1+x^2)}\right) \nn\\
 && \quad \cdot \e^{-\xi^2/(2a(1+x^2))} \label{eqn:IT12} \\
\xilarge -2^{-3/2}\sqrt{\pi}a^{-5/2} \xi^2
 \int_0^1 dx \cx{1/2}{-5/2}\e^{-\xi^2/(2a(1+x^2))} \\
%%%
I_\chi &=& 2^{-3/2}\sqrt{\pi} a^{-1/2}\int_0^1 dx \cx{1/2}{-1/2}
 \e^{-\xi^2/(2a(1+x^2))} \label{eqn:IC12}\\
I_f &=& \half \Gamma\left( {3\over 4}\right) a^{-3/4}
 \Fiia{{3\over 4}}{\half} \\
\xilarge -{\sqrt{\pi} \over 2\sqrt{2}}|\xi|^{-3/2} 
\eqaend
%%%%%%%%%%%%%%%%5
\parag{$d=1$ case }
%%%%%%%%%%%%%%%%%%
\eqabegin
I_T &=& {1\over 2a^2}\int_0^1dx (1-x^2)(1+x^2)^{-2}
 \Fiialpha{2}{\half} \\
\xilarge {52\over 35} |\xi|^{-4} \\
%%%%%
I_\chi &=& {1\over 4a} \int_0^1dx (1-x^2)(1+x^2)^{-1}
 \Fiialpha{1}{\half} \\ 
\xilarge -{1\over 5}|\xi|^{-2} \\
I_f &=& {1\over 2a} \Fiia{1}{\half} \\
\xilarge -|\xi|^{-2} 
\eqaend
%%%%%%%%%%%%%%%%%%%%%%%%%%%%%%%%%%%%%%%

%%%%%%%%%%%%%%%%%%%%%%%%%%%%%%%%%%%%%%%%%%%
%    figures 
%%%%%%%%%%%%%%%%%%%%%%%%%%%%%%%%%%%%%%%%%%
\newpage
\begin{center}
{\Large\bf Figure Captions }\\
\end{center}
Fig.1a : A plot of $ I_T ( \xi ) $  for $ d = -1/2 $ and $ a = 0.005 $. 
        The shape is very nearly a steeply peaked Gaussian.\\
\\
Fig.1b : A plot of $ I_{\chi} ( \xi ) $ up to a constant  for $ d = -1/2 $
        and $ a = 0.005 $.
        The asymptotic behavior for large $ \mid \xi \mid $ is linear.\\  
\\
Fig.1c : The line of curvature singurarity ( solid line ) for $ d = -1/2 $ 
        produced by a left-going smeared shock wave along 
        $ \xi^+ = \xi^+_0 $ ( dotted line ). The dot-dashed line represents 
        the event horizon and the space-time quickly approaches the linear 
        dilaton vacuum to the left of $ \xi^+ = \xi^+_0 $. \\
\\
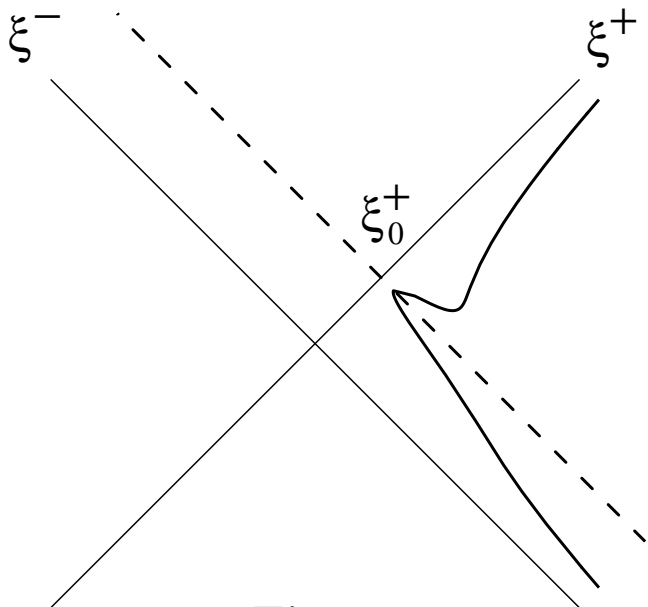
Fig.2 : The line of curvature singularity ( solid line ) for $ d = 1/2 $
        produced by a left-going smeared shock wave along 
        $ \xi^+ = \xi^+_0 $ ( dotted line ). Note the appearance of naked 
        singularity.\\
\newpage

\setlength{\oddsidemargin}{0cm}
\setlength{\baselineskip}{7mm}

\hspace{0.9in}
%\vspace{0.8in}
\epsfbox{figure1a.ai}\\

%\vspace{0.5in}
\hspace{2.6in}{\Large\bf Fig.1a}

\vspace{1in}

\hspace{0.9in}
\epsfbox{figure1b.ai}

%\vspace{0.5in}
\hspace{2.6in}{\Large\bf Fig.1b}

\vspace{-1.5in}
\hspace{0.1in}
%\epsfbox[93 300 495 400]{figure1c.ai}
\epsfbox{figure1c.ai}

\vspace*{-1.5in} 
\hspace*{2.55in} {\Large\bf Fig.1c}

%\vspace{-0.8in}
\hspace{0.1in}
\epsfbox{figure2.ai}

\vspace*{-1.5in}
 \hspace*{2.55in} {\Large\bf Fig.2}
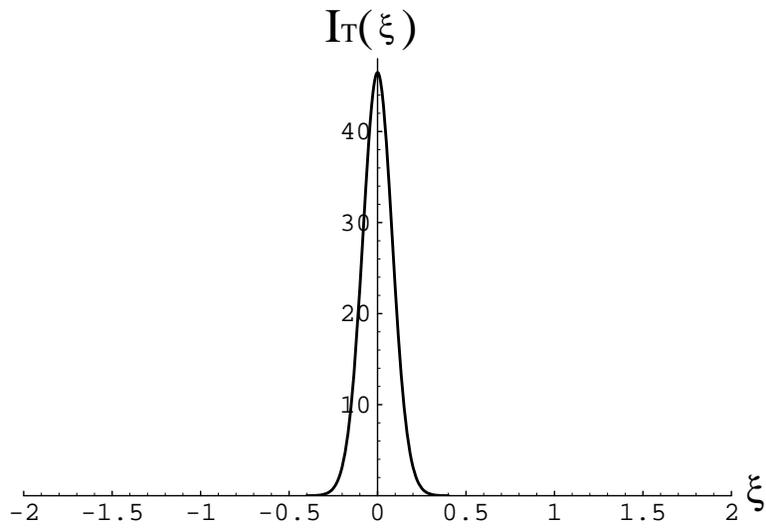
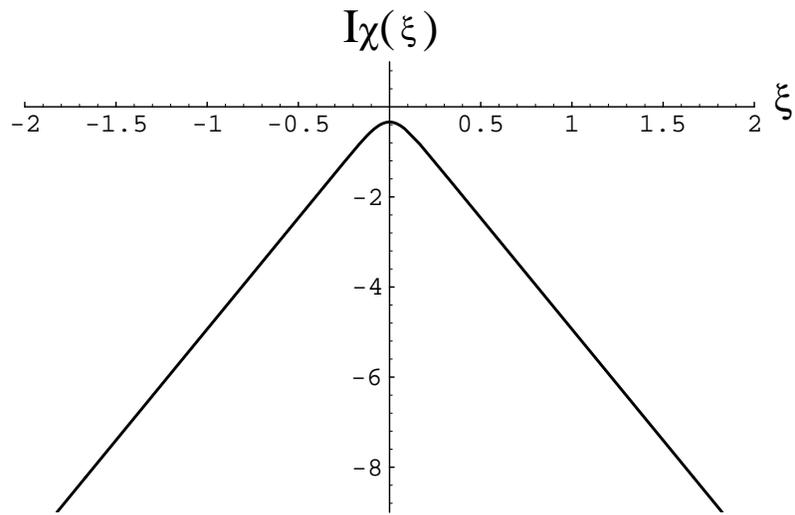
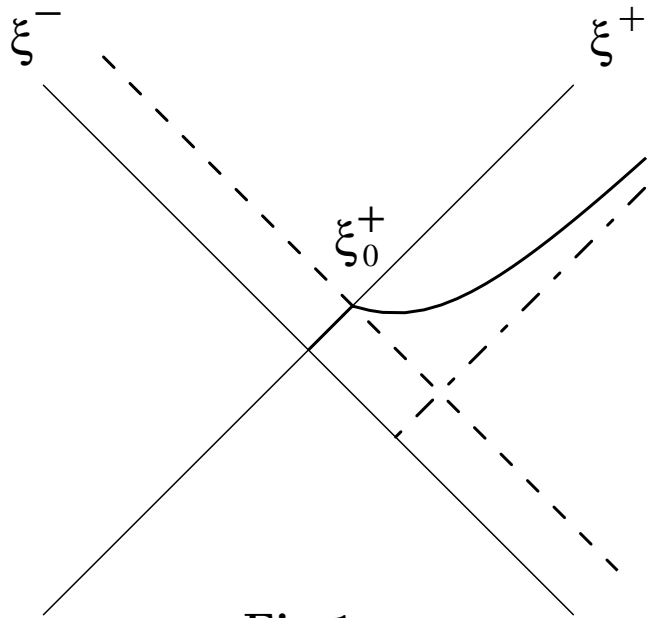
\end{document}